\documentclass{emulateapj}
\usepackage{graphicx}
\usepackage[usenames]{color}
\slugcomment{Accepted to The Astrophysical Journal,  Nov 17, 2014}

\newcommand \spitzer {{\it Spitzer}}

\newcommand \msun  {M$_{\odot}$}

\newcommand	      \myr              {M$_{\odot}$~yr$^{-1}$}

\newcommand      \cc       {cm$^{-3}$}
\newcommand \mg      {$\overline{m}_g$}
\newcommand \md      {$\overline{m}_d$}
\newcommand \mge      {\overline{m}_g}
\newcommand \mde      {\overline{m}_d}
\newcommand \taud      {$\tau_d$}
\begin{document}

%\received{}
%\revised{}
%\accepted{}

\shortauthors{TEMIM ET AL.}

\shorttitle{DUST FORMATION AND DESTRUCTION} 

\title{DUST DESTRUCTION RATES AND LIFETIMES IN THE MAGELLANIC CLOUDS}

\author{TEA TEMIM\altaffilmark{1,2}, ELI DWEK\altaffilmark{1}, KIRILL TCHERNYSHYOV\altaffilmark{3}, MARTHA L. BOYER\altaffilmark{1,4}, MARGARET MEIXNER\altaffilmark{3,5}, CHRISTA GALL\altaffilmark{6,7}, JULIA ROMAN-DUVAL\altaffilmark{5}}

\altaffiltext{1}{Observational Cosmology Lab, Code 665, NASA Goddard Space Flight Center, Greenbelt, MD 20771, USA}
\altaffiltext{2}{CRESST, University of Maryland-College Park, College Park, MD 20742, USA}
\altaffiltext{3}{Department of Physics and Astronomy, The Johns Hopkins University, 366 Bloomberg Center, 3400 North Charles Street, Baltimore, MD 21218, USA}
\altaffiltext{4}{Oak Ridge Associated Universities (ORAU), Oak Ridge, TN  37831, USA; tea.temim@nasa.gov}
\altaffiltext{5}{Space Telescope Science Institute, 3700 San Martin Drive, Baltimore, MD 21218, USA}
\altaffiltext{6}{Department of Physics and Astronomy, Aarhus University, Ny Munkegade 120, DK-8000 Aarhus C, Denmark}
\altaffiltext{7}{Dark Cosmology Centre, Niels Bohr Institute, University of Copenhagen, Juliane Maries Vej 30, DK-2100 Copenhagen ¯, Denmark}

\begin{abstract}

The nature, composition, abundance, and size distribution of dust in galaxies is determined by the rate at which it is created in the different stellar sources and destroyed by interstellar shocks. Because of their extensive wavelength coverage, proximity, and nearly face-on geometry, the Magellanic Clouds (MCs) provide a unique opportunity to study these processes in great detail. In this paper we use the complete sample of supernova remnants (SNRs) in the MCs to calculate the lifetime and destruction efficiencies of silicate and carbon dust in these galaxies. We find dust lifetimes of $22\pm13$ Myr ($30\pm17$ Myr) for silicate (carbon) grains in the LMC, and $54\pm32$ Myr ($72\pm43$ Myr) for silicate (carbon) grains in the SMC. 
%The corresponding dust destruction rates are $(2.3\pm1.3)\times10^{-2}$ \myr\ and $(3.6\pm1.8)\times10^{-3}$ \myr\ for the LMC and SMC, respectively. 
The significantly shorter lifetimes in the MCs, as compared to the Milky Way, are explained as the combined effect of their lower total dust mass, and the fact that the dust-destroying isolated SNe in the MCs seem to be preferentially occurring in regions with higher than average dust-to-gas (D2G) mass ratios. We also calculate the supernova rate and the current star formation rate in the MCs, and use them to derive maximum dust injection rates by asymptotic giant branch (AGB) stars and core collapse supernovae (CCSNe). We find that the injection rates are an order of magnitude lower than the dust destruction rates by the SNRs. This supports the conclusion that, unless the dust destruction rates have been considerably overestimated, most of the dust must be reconstituted from surviving grains in dense molecular clouds. 
More generally, we also discuss the dependence of the dust destruction rate on the local D2G mass ratio and the ambient gas density and metallicity, as well as the application of our results to other galaxies and dust evolution models.
\end{abstract}


%=========================================
\section{INTRODUCTION} \label{intro}
%=========================================

The evolution of dust grains in galaxies is driven by their formation rate in the different stellar sources, their processing by supernova (SN) blast waves, and by their reconstitution in dense interstellar clouds. Dust formation sites include the ejecta of core collapse and Type~Ia supernovae (CCSNe, SNIa, respectively), novae, and mass outflows from evolved stars in the Asymptotic Giant Branch (AGB) phase of their evolution and Wolf-Rayet (WR) stars.
An important issue is the nature of interstellar dust. Is it entirely made of refractory elements that thermally condensed in stellar ejecta, or does it also contain heavy elements that accreted onto these refractory cores in the dense phases of the interstellar medium (ISM)?

%==================================================================
\begin{figure*}
\epsscale{0.55} \plotone{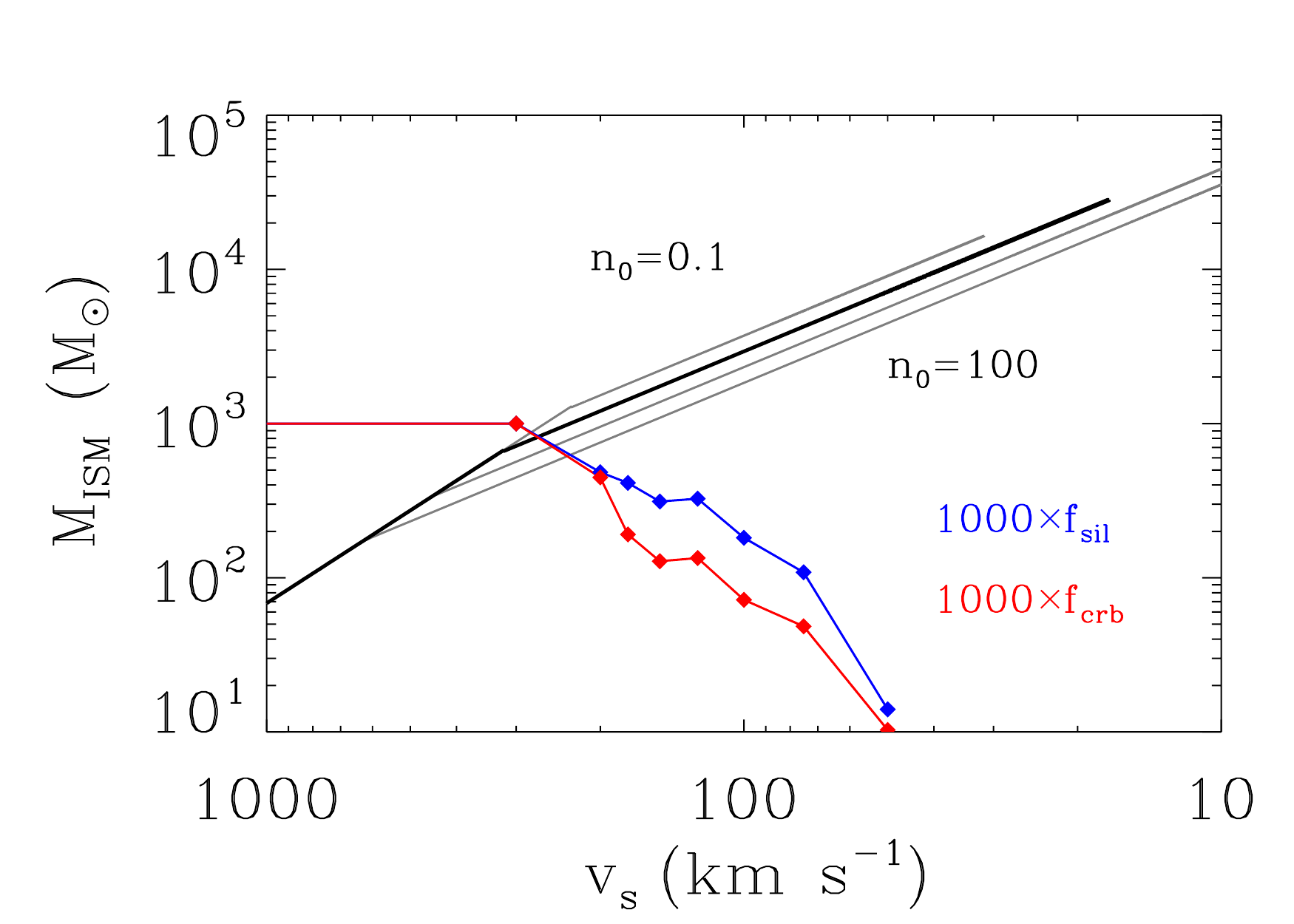}
\epsscale{0.55} \plotone{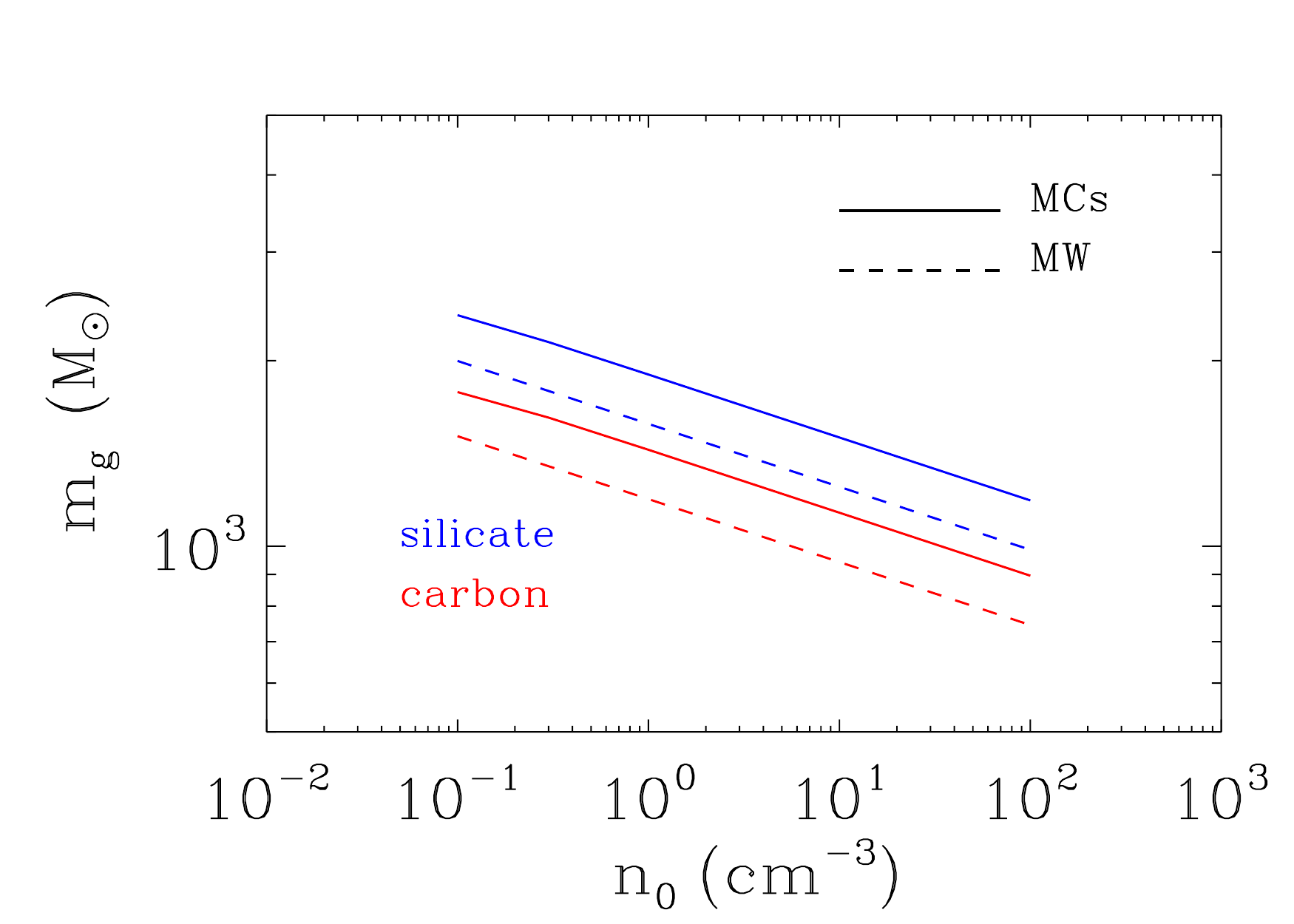}
\caption{\label{snr_mass} {\bf Left panel}: Mass of ISM material swept up by a pressure driven SNR, as a function of shock velocity. The grey curves represent swept-up masses for different ISM densities ranging from $\rm n_0=0.1-100\: cm^{-3}$, while the black curve represents $\rm n_0=1\: cm^{-3}$. The blue (red) curves represent grain destruction efficiencies for silicate (carbon) grains, multiplied by a factor of 1000 for display purposes. The efficiencies were taken from Table~4 of \citet{jones96}, with additional
calculations for intermediate velocities provided by Slavin (private communication). {\bf Right panel}: The total effective swept-up mass $m_g$ as a function of ISM density for silicate (blue) and carbon (red) grains. The total mass $m_g$ at each density is the convolution of the swept-up ISM mass $M_{ISM}$ and the grain destruction efficiency $f_d(v_s)$, integrated over the shock velocity. It can be seen that $m_g$ is only weakly dependent on the ISM density ($\propto n^{-0.107}$). The solid line represents $m_g$ values for the MCs, while the dashed line represents values for the MW, which is lower due to its higher metallicity.}
\end{figure*}
%==================================================================

These questions were first raised with the realization that the interstellar elemental depletion pattern correlated with the condensation temperature \citep{field74}, suggesting that thermal condensation in the stellar sources drives the evolution of dust. An equally good correlation of the elemental depletion pattern with the first ionization potential was suggested by \cite{snow75} as evidence that dust was primarily grown by accretion in the ISM. This issue was first quantitatively addressed by \cite{dwek80b} who examined the balance between the formation rates of dust in CCSNe and AGB stars and their destruction rate by SN remnants (SNRs) in the framework of a chemical evolution model. They noted that any deficiency between the rate of dust formation and destruction may necessitate grain growth by accretion in the ISM, in order to explain the abundance of dust in the diffuse ISM, as inferred from the elemental depletion pattern. This conclusion was confirmed by detailed calculations of the grain destruction rates \citep{jones96,slavin14}, and more detailed chemical evolution models \citep{dwek98, tielens98, zhukovska08a, calura10} for the solar neighborhood. 

The need for the ISM accretion to explain the inferred dust abundance in local and high-redshift galaxies was also confirmed in chemical evolution models \citep{valiante09, gall11a, dwek11a, dwek11b}. The problem is primarily the consequence of the fact that CCSNe are net destroyers of dust, that is, they destroy more dust in the remnant phase of their evolution than they form in the ejecta. When this difference is not made up by dust production in AGB stars, the ``missing" dust must be grown onto the surviving refractory grain cores in the ISM. Only in very high redshift galaxies, in which the dust to gas mass ratio is $\lesssim 10^{-4}$, are CCSNe  net producers of interstellar dust, alleviating the need to reconstitute the dust in the ISM \citep{dwek14}.

The question of the imbalance between the dust destruction and production rates can be uniquely addressed by studies of the Large and Small Magellanic Clouds, LMC and SMC, respectively.
These galaxies are the best astrophysical laboratories to study the 
lifecycle of dust in galaxies. Their proximity (50
kpc, e.g. \citet{schaefer08} and 62 kpc, \citet{szewczyk09}) permits detailed
studies of individual stars and stellar populations. These can be used to derive their star formation history \citep{harris04, harris09}, and dust production rates from carbon- and oxygen-rich stars \citep{srinivasan09,matsuura09, boyer10,riebel12,boyer12,matsuura13}. Their almost face-on geometry reveals a fairly complete sample of SNRs \citep{badenes10}.  Most importantly, recent far infrared (IR) observations of the MCs with the \textit{Herschel} Space Observatory \citep{meixner13}, allow us to observationally determine the properties of the environment into which each SNR is expanding, and therefore determine the most reliable current rate of grain destruction in the MCs. 

In this paper we use the observationally determined ISM density and dust-to-gas (D2G) mass ratio around a nearly complete sample of SNRs in the MCs to calculate the global rate of grain destruction and the corresponding dust lifetimes. For comparison, we also calculate a maximum rate of grain formation in the quiescent outflows of AGB stars and the explosively ejected material in CCSN events. The main purpose is to determine if the balance between the injection and destruction of dust in the MCs is consistent with currently observed dust emission.

The paper is organized as follows. In \S\ref{equations} we present the general equations for calculating the grain destruction rate by SNRs and identify the parameters that determine this rate. In \S\ref{results} we describe how we derive each of the required parameters, and use the H\small{I} gas and {\it Herschel} Space Observatory IR images to calculate the ISM density and dust-to-gas (D2G) mass ratio of the medium into which each of the SNRs expand, and the masses of dust destroyed during the evolution of the SNRs. In \S\ref{lifetimes}, we give the corresponding dust destruction rates and lifetimes and discuss how they compare to those of the Milky Way. In \S\ref{application}, we discuss how our results can be applied to other galaxies and dust evolution models. In \S\ref{production},  we derive upper limits on the dust formation rates in AGB and CCSNe, and compare them to the dust destruction rates that we derive in this work.  Our major goal is to determine if there is a discrepancy in the balance between the dust formation and destruction rates in the MCs that may require an additional source of dust. The astrophysical implication of our results are discussed in \S\ref{implications}.

%==================================================================
%------ \small{HI} surface density map
\begin{figure}
\epsscale{1.20} \plotone{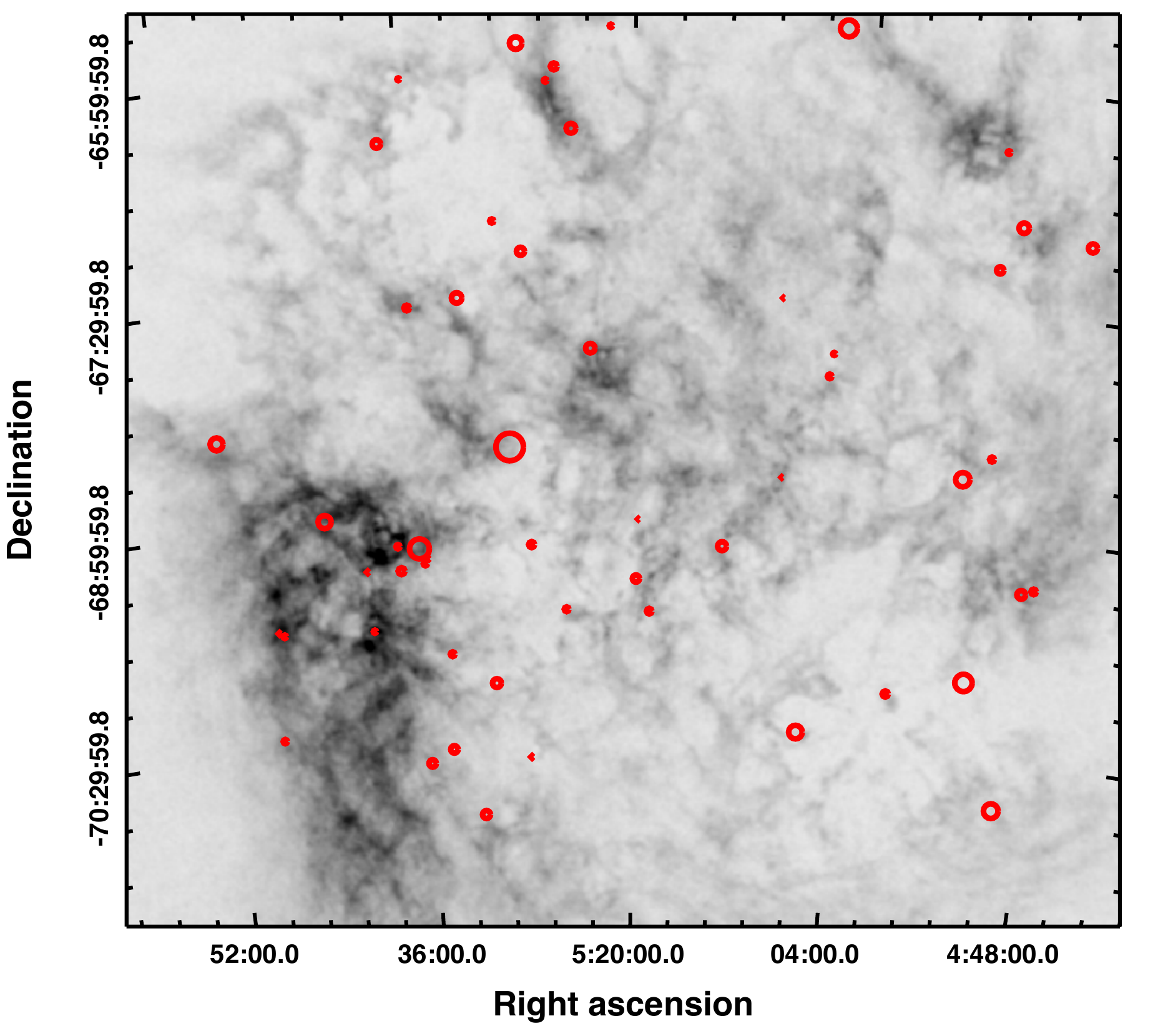}
\epsscale{1.20} \plotone{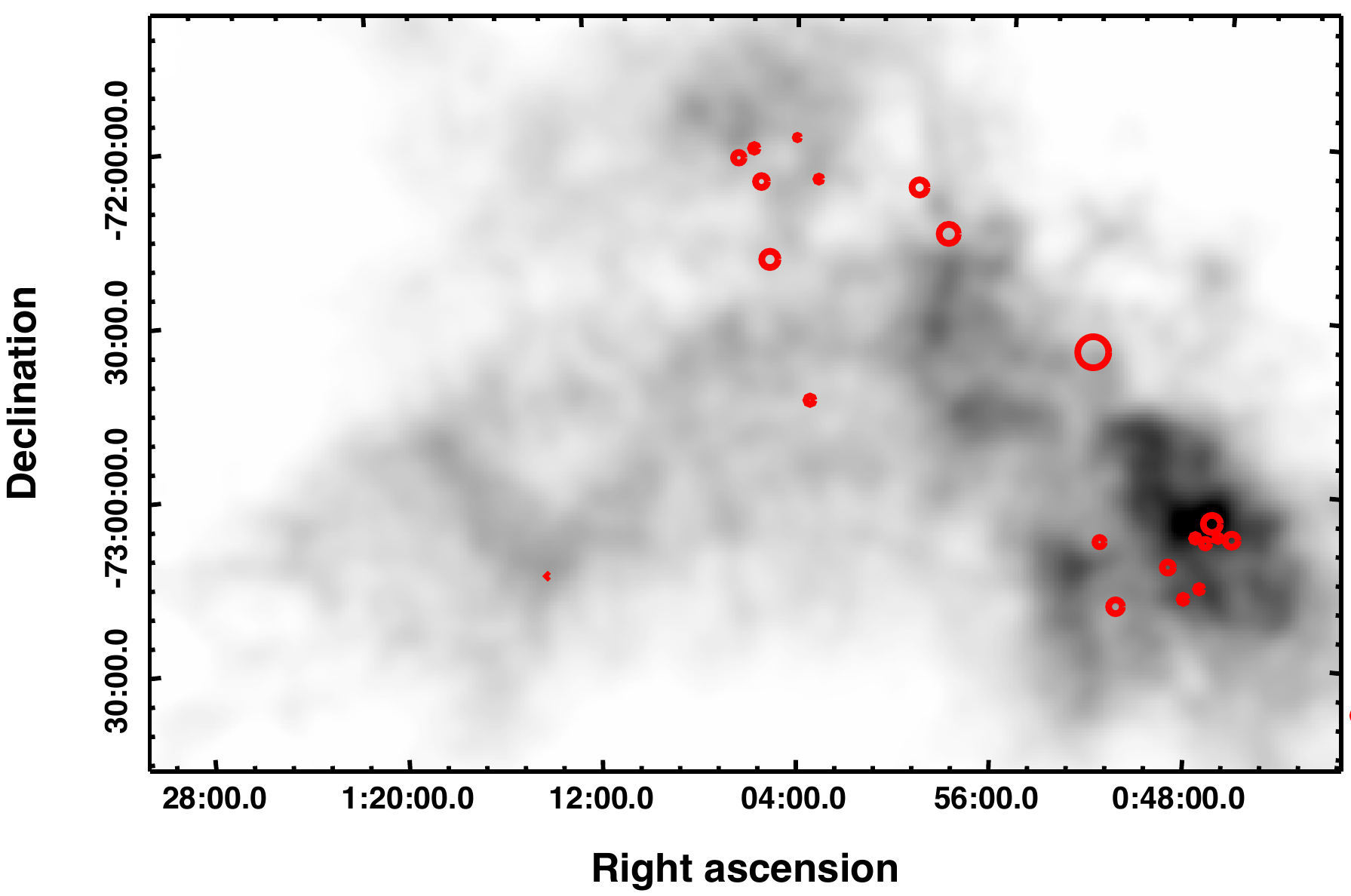}
\caption{\label{maps}HI surface density maps of the LMC \citep{kim99} and SMC \citep{stanimirovic99} with the positions of observed SNRs overlaid as red circles. The size of the circles represents the measured SNR sizes as listed in \citet{badenes10} and \citet{lakicevic14}.}
\end{figure}

%==================================================================

%==================================================================
%\section{THE RATE OF GRAIN DESTRUCTION BY \\SUPERNOVA REMNANTS} 
\section{GENERAL EQUATIONS}\label{equations}
%==================================================================

%-------------------------------
%\subsection{General Equations}\label{equations}
%-------------------------------
Isolated SNRs expanding into the ISM destroy interstellar grains by thermal-kinetic sputtering and vaporizing grain-grain collisions.
There are several distinct lines of evidence for grain destruction and processing in shocks: (1) X-ray observations showing changes in the elemental abundances and ionization structure of heavy elements in the postshock flow \citep{vancura94,raymond13} ; (2) UV observations showing changes in the depth of the 2200 \AA\ extinction feature and in the slope of the extinction across the shock \citep{seab83}; (3) analysis of IR observations that show changes in the grain size distribution before and after the shock \citep{arendt10, sankrit10}, (4) significantly lower than average D2G mass ratio in LMC SNRs  \citep{williams06a, borkowski06, williams11a}; and (5) differences in dust density maps along the line of sight to the LMC SNRs \citep{lakicevic14}.

The dust mass destroyed by a single SNR, $m_d$, can be expressed by \citep{jones94, dwek07b}:
\begin{equation}
\label{meff}
m_d = Z_d\, m_g = Z_d\ \int_{v_i}^{v_f}\ f_d(v_s)\, \left|{dM_{\rm ISM}\over dv_s}\right|\, dv_s.
\end{equation}
where $Z_d$ is the D2G mass ratio of the local ISM into which the remnant expands, $m_g$ is the effective mass of ISM gas that is completely cleared of dust by a single SNR, $dM_{\rm ISM}/ dv_s$ is the rate at which the ISM is swept up by the SNR as a function of shock velocity $v_s$, and $f_d(v_s)$ is the fraction of dust that is destroyed as a function of shock velocity.  For example, models of the pre- and postshock \textit{Spitzer} Infrared Spectrograph (IRS) spectra of Puppis~A \citep{arendt10} and the Cygnus Loop \citep{sankrit10} show that $\sim$ 25--30\% of the grains are destroyed in $\sim 500$~km~s$^{-1}$ shocks.
The limits of the integral extend from the initial velocity of the remnant, $v_i$ to $v_f$, its final velocity when the ejecta reaches the random velocities of the ISM. 

For our analysis, we compute the SNR evolution and $dM_{ISM}/ dv_s$ values using the pressure-driven snowplow model of \citet{cioffi88}. The results are shown in Figure~\ref{snr_mass} (left panel), where the gray and black curves represent the total mass of ISM material swept up by an SNR as a function of shock velocity for ISM densities ranging from $\rm n_0=0.1-100\: cm^{-3}$, and an explosion energy $E_0=10^{51}$ erg.

 The right panel of Figure~\ref{snr_mass} shows the total effective swept-up mass $m_g$ as a function of ISM density for silicate and carbon grains, using the grain destruction efficiencies from \citep[][,Table 4]{jones96}, with additional calculations for intermediate velocities provided by Slavin (2014, private communication, in prep). The total mass $m_g$ is simply the convolution of the swept-up ISM mass $M_{ISM}$ and the grain destruction efficiency $f_d(v_s)$ in the left panel of Figure~\ref{snr_mass}, integrated over the shock velocity.  The weak dependence of $m_g$ on the ISM density implies that the primary factor that determines the dust mass destroyed by an SNR, $m_d$, is the D2G mass ratio in the preshocked ISM.

The dust lifetime \taud\ at the current epoch can be written as:
%-----------------
\begin{equation}
\label{taud}
\tau_d = {M_d \over \mde R_{SN}}
\end{equation}
where $M_d$ is the total dust mass in the galaxy, $R_{SN}$ is the total (CCSN + Type~Ia) supernova rate, and \md=$\sum{m_d}/N_{SNR}$, the mass of refractory elements initially locked up in dust, averaged over the total number of SNRs in the galaxy $N_{SNR}$. In rest of the paper, a bar above a symbol will indicate an appropriate average value over the total number of SNRs in the galaxy.

The current grain destruction rate, $dM_d/dt$, is given by:
%-----------------
\begin{equation}
\label{dmd_dt}
{dM_d \over dt} = {M_d\over \tau_d} = \mde R_{SN}.
\end{equation}

For a given explosion energy, $E_0$,  the rate of grain destruction depends on: (1) $f_d(v_s)$, the grain destruction efficiency; (2)  $n_0$ and $Z_d$, the density and D2G mass ratio, respectively, of the preshocked ISM into which the SNR is expanding; and (3) $R_{SN}$, the supernova rate. To determine the dust lifetime, we also need to know the total mass of dust in the galaxies $M_d(t)$. In this work, we use observationally determined values for $M_d$, and $n_0$ and $Z_d$ for all confirmed SNRs, to calculate the value of the average destroyed dust mass \md, the dust lifetime $\tau_d$, and the global dust destruction rate $dM_d(t)/dt$ for the Magellanic Clouds. Below, we outline the derivation method and results for each of the parameters.

%==================================
%------ histogram of density distribution
\begin{figure*}
\epsscale{0.53} \plotone{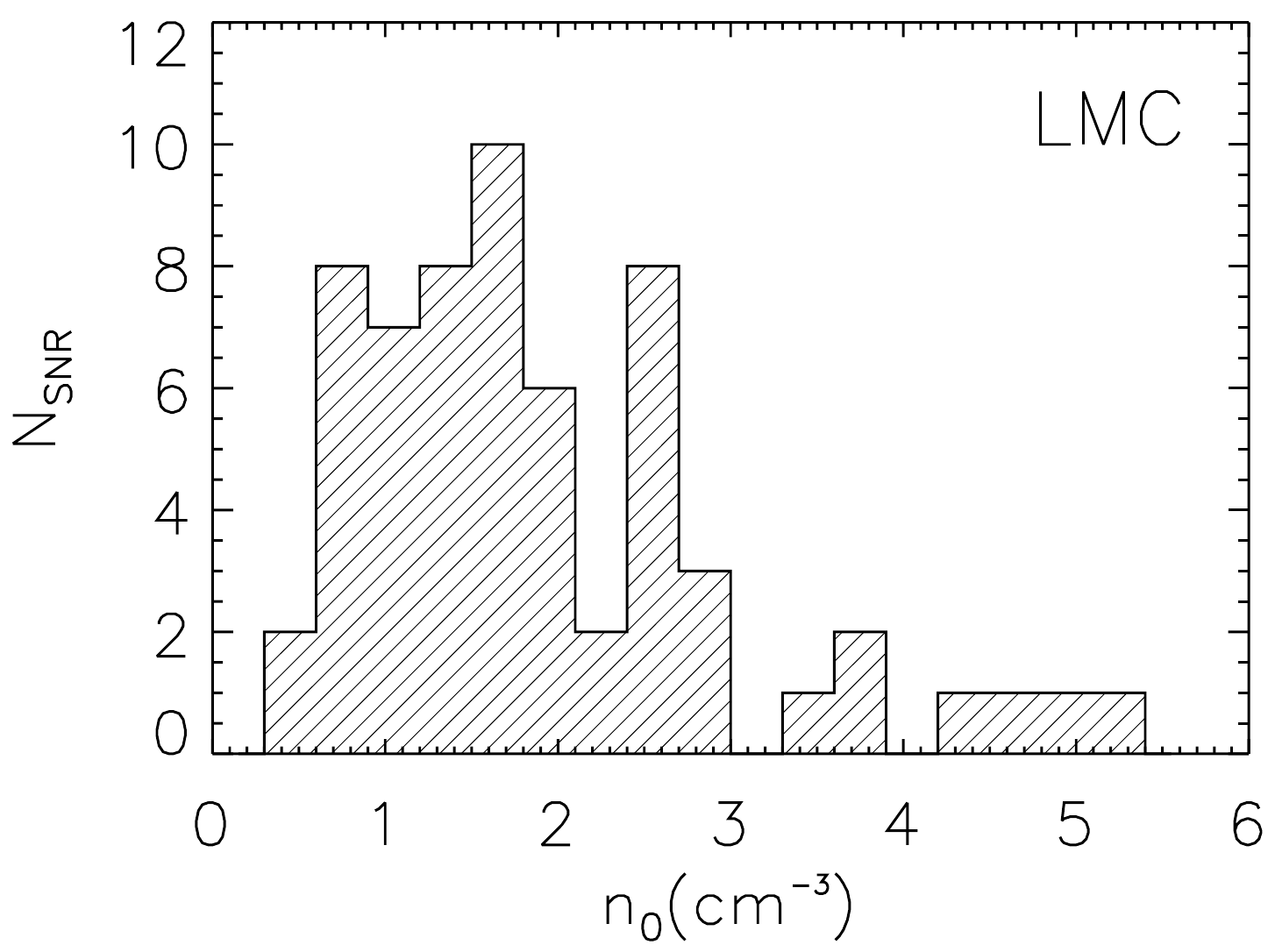}
\epsscale{0.51} \plotone{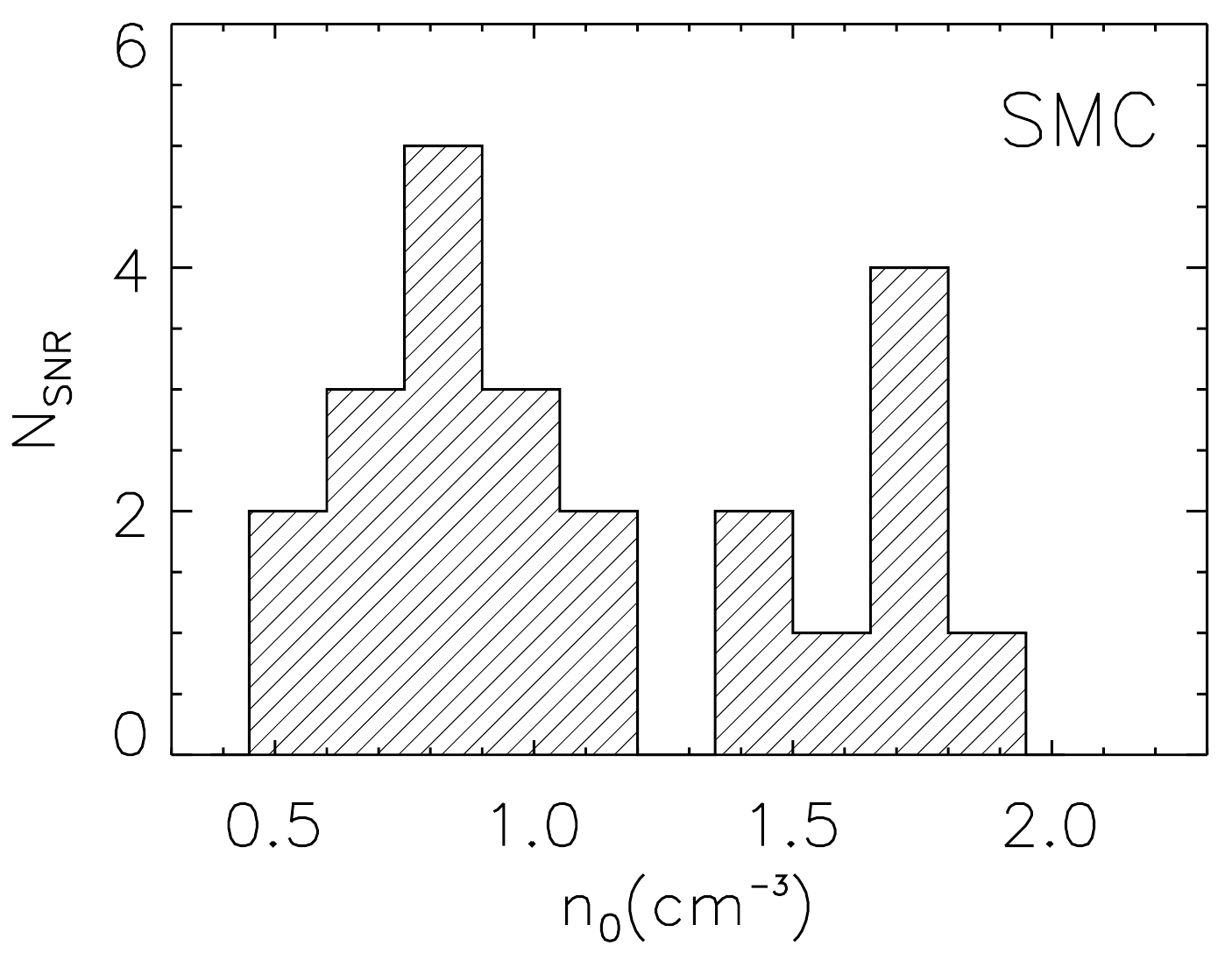}
\epsscale{0.53} \plotone{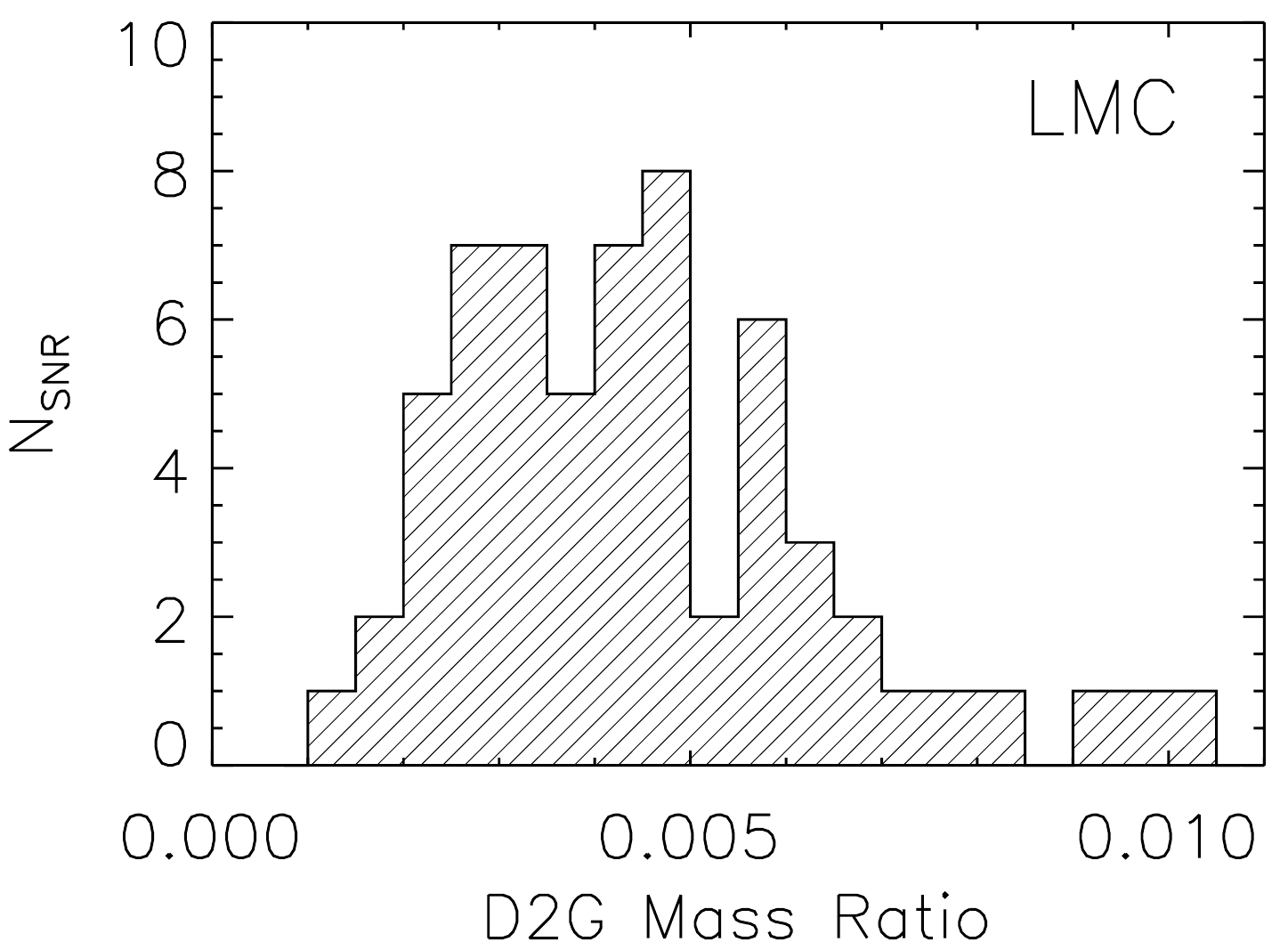}
\epsscale{0.51} \plotone{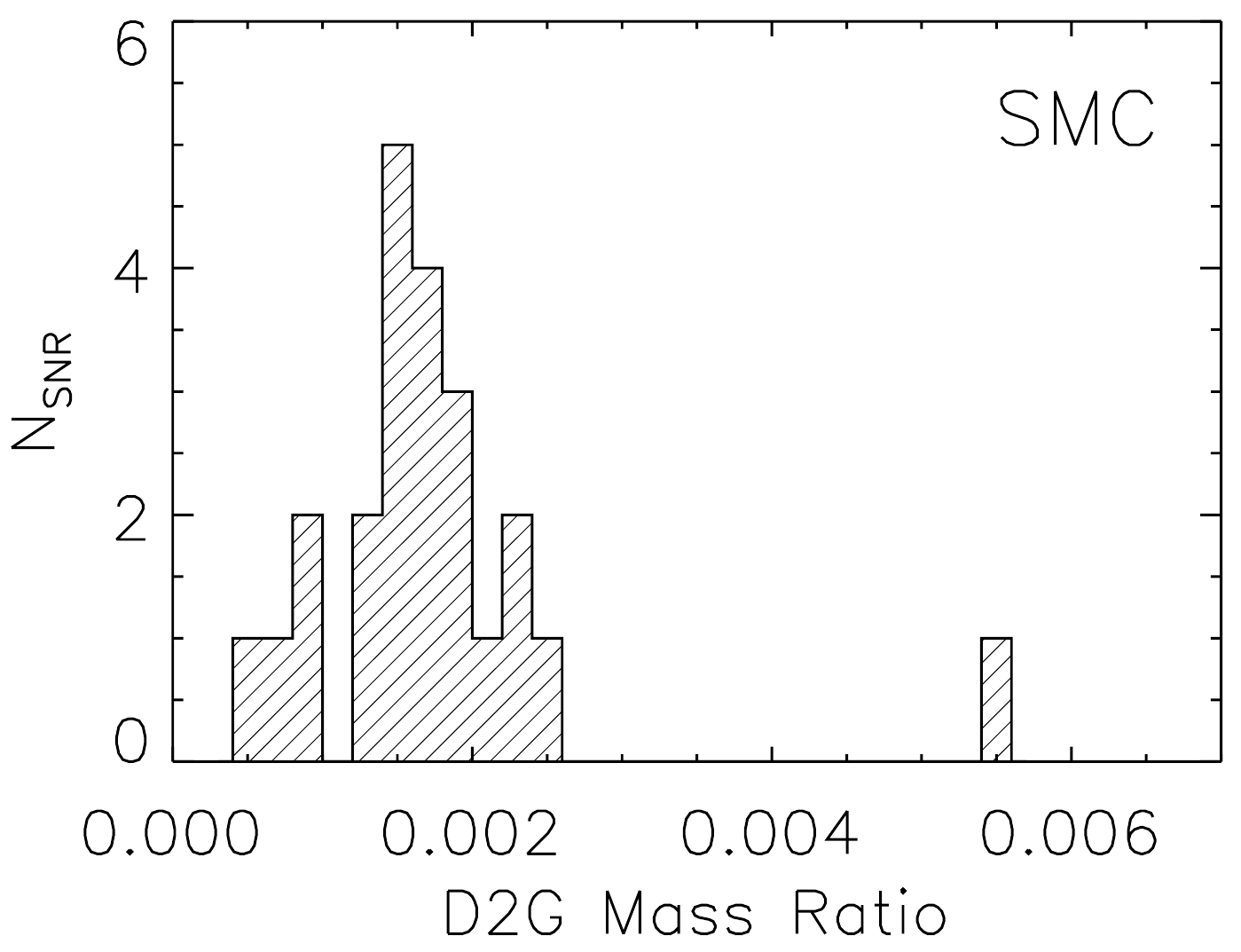}
\caption{The histograms show the distribution of ambient gas densities (assuming a disk thickness of 0.4 kpc for the LMC and 2.0 kpc for the SMC) and D2G mass ratios for regions surrounding each of the SNRs.\label{nm_g2d}}
\end{figure*}
%==================================

%==================================
\section{ANALYSIS \& RESULTS}\label{results}
%==================================

%--------------------------------------------
\subsection{SNR Sample}\label{sample}
%--------------------------------------------

For our analysis, we used a list of all known SNRs in the LMC and SMC \citep[e.g.][]{chu88,williams99,filipovic05,blair06,badenes10,seok13,maggi14,lakicevic14}. \citet{badenes10} carefully assessed the completeness of the SNR sample, and concluded that the list is a fairly complete sample that should not be missing a large number of objects. Following their study, four new X-ray selected SNRs were confirmed in the LMC by \citet{maggi14}.
Tables \ref{snrs_lmc} and \ref{snrs_smc} list the names, coordinates, and diameters of all confirmed SNRs in the MCs. The list contains 61 SNRs in the LMC and 23 SNRs in the SMC.
SNR diameters were measured from X-ray observations, when available, and from the best available alternative, when not.
The spatial distribution of SNRs, along with their spatial sizes are shown in Figure~\ref{maps}. The positions and sizes are indicated by the red circles, and overlaid onto the H{\small{I}} surface density maps of the LMC \citep{kim99} and SMC \citep{stanimirovic99}.

%--------------------------------------------
\subsection{The ISM Density}\label{density}
%--------------------------------------------
The ISM density determines the rate at which the ISM mass is swept up at any given shock velocity, and hence overall the grain destruction efficiency by SNRs. 
To derive the pre-shock density of the ISM surrounding each SNR in the LMC and SMC, we assumed that they are expanding into a a neutral hydrogen medium. We therefore extracted \small{HI} column density values in annuli at each SNR position. The \small{HI} maps were convolved to the \textit{Herschel} SPIRE 500 \micron\ image (14\arcsec/pixel) for consistency, and values were extracted from square annuli centered at each SNR, with the inner square side equal to 1 $\times$ SNR diameter, and the outer square side equal to 4 $\times$ SNR diameter. For SNRs that were smaller than one resolution element, values were extracted from a 9 $\times$ 9 pixel region, minus the central pixel. The average \small{HI} column density $N_H$ from each of these regions for both LMC and SMC remnants is listed in Tables \ref{snrs_lmc} and \ref{snrs_smc}.

In order to obtain an absolute gas density into which the SNRs are expanding, we need to divide $N_H$ by the thickness of the gas disks of the LMC and SMC. We adopt an LMC gas disk thickness of 0.4 kpc \citep{kim99}, and an SMC gas disk thickness of 2.0 kpc \citep{stanimirovic04}, both based on measurements from \small{HI} observations of the MCs. In order to explore the dependence of our results on the choice of disk thickness, we computed dust destruction rates and lifetimes for a range of disk thickness values, 0.05--1.6 kpc for the LMC, and 1.0--4.0 kpc for the SMC. The upper limits for the disk thickness values were chosen based on the LMC and SMC depth estimates from observations of Cepheids and RR Lyrae stars \citep{haschke12a,haschke12b}.

Figure \ref{nm_g2d} (top) shows  the distribution of absolute densities for both the LMC and SMC. These densities are also listed in Tables \ref{snrs_lmc} and \ref{snrs_smc}. The mean density $\rm n_0$ around the SNR sample in the LMC and SMC as function of disk thickness is listed in Table \ref{snrs_avg}. 

%==============================================================
%------ SED fits
\begin{figure*}
\epsscale{0.55} \plotone{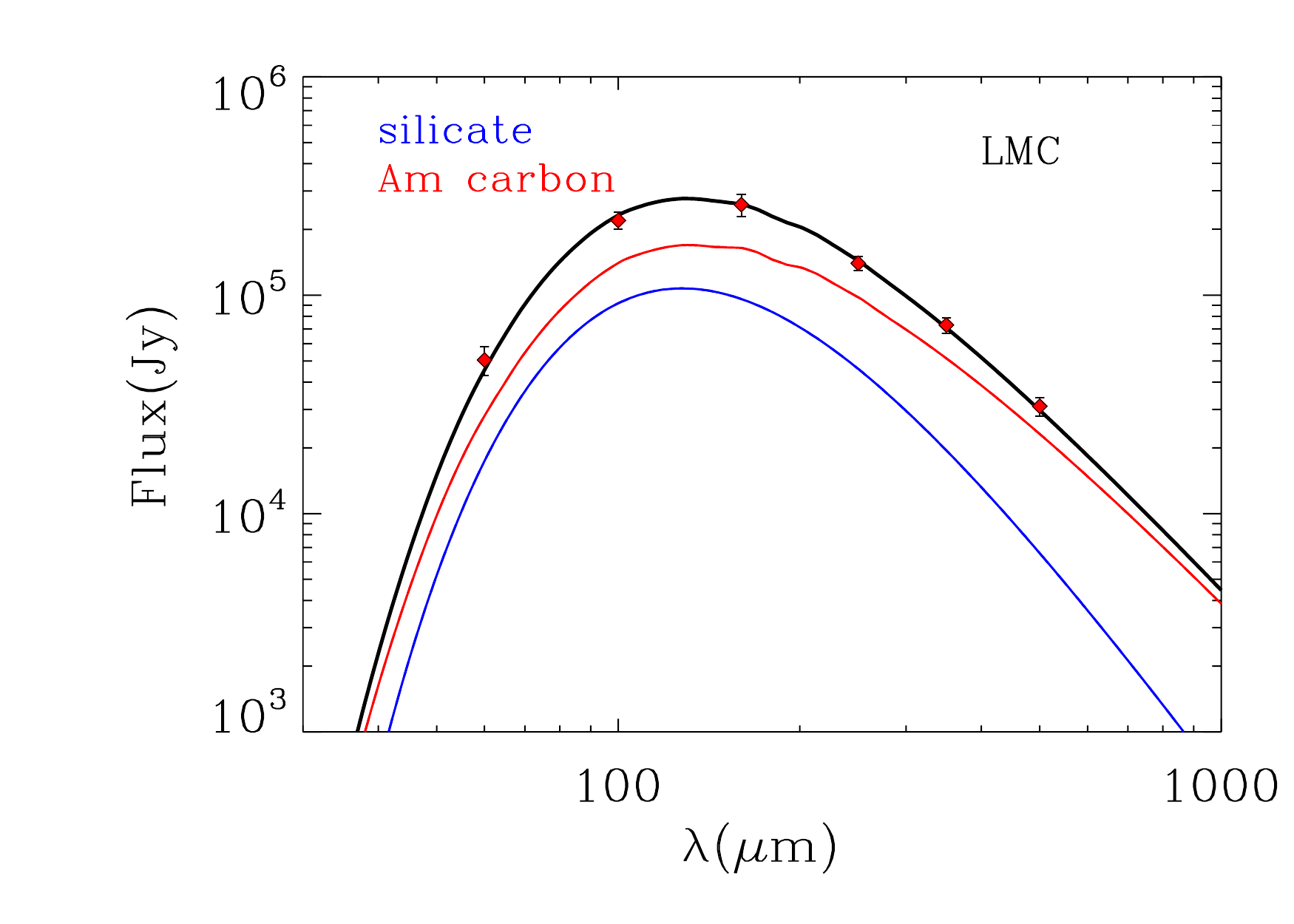}
\epsscale{0.55} \plotone{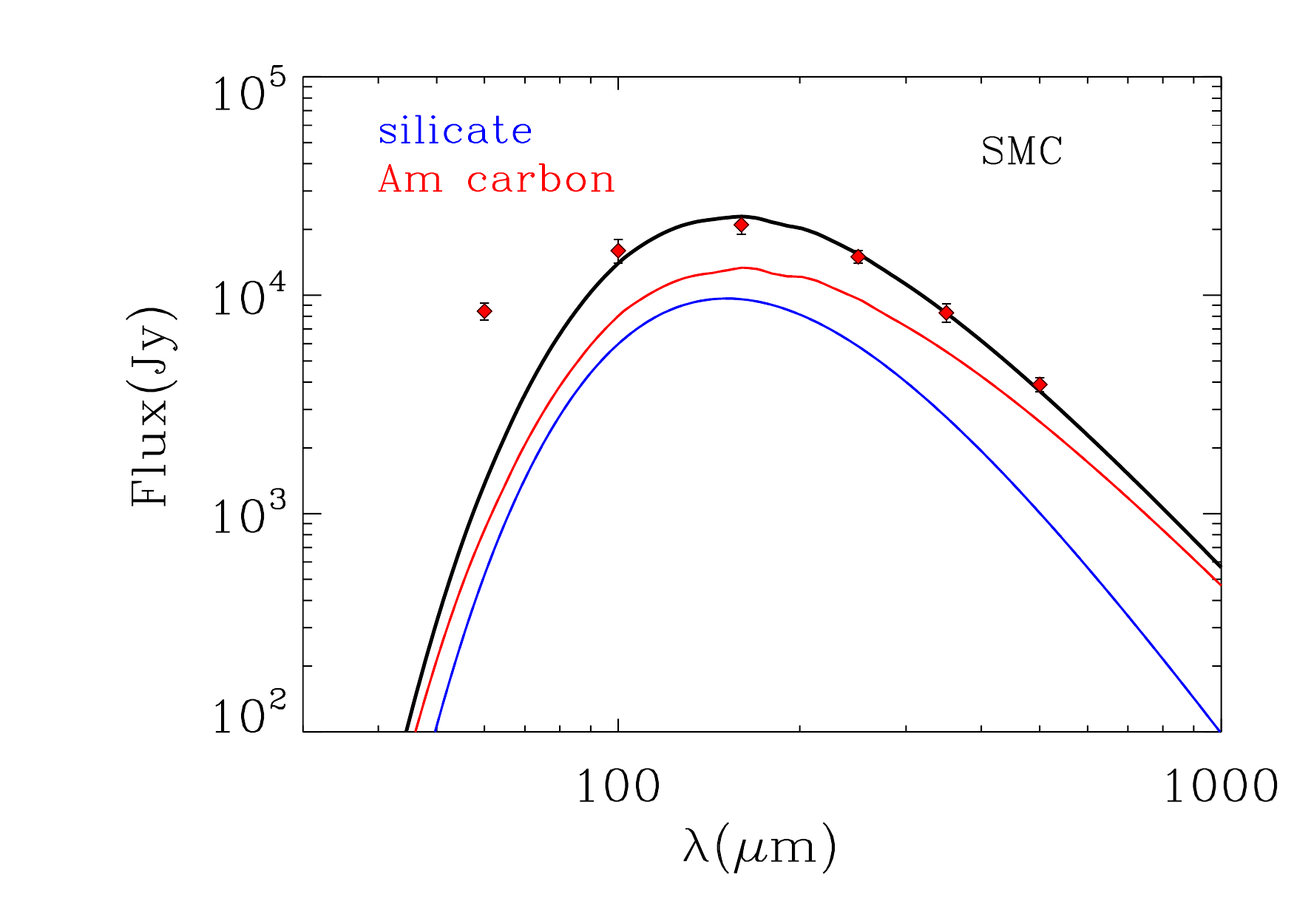}
\caption{\label{sed_fits}Constrained two component fits to the SEDs of the LMC and SMC. See Section \ref{tot_mass} for details and Table \ref{params} for best fit masses and temperatures.}
\end{figure*}
%==============================================================

%--------------------------------------------
\subsection{Total Gas and Dust Masses in the \\Magellanic Clouds}\label{tot_mass}
%--------------------------------------------

A map of the dust mass distribution and the total mass of dust in the LMC were recently derived by \cite{gordon14}, using a  parametrized representation of the dust properties. 
Since we  are interested in determining the separate contributions from silicate and carbon dust, which were not computed in previous works, we recalculated the dust masses using the dust optical constants from \citep{li02} for silicate, and \citep{zubko96} for amorphous carbon dust. 
The 70--500 \micron\ infrared fluxes were taken with the \textit{Herschel} PACS and SPIRE instruments, and are presented in  \cite{meixner13}.

To fit the dust models to the data, we adopted two physical constraints on the dust. The first uses the LMC and SMC elemental abundances \citep[][and references therein]{tchernyshyov14} to constrain the relative silicate-to-carbon dust mass ratio to 2.9 for the LMC, and 4.0 for the SMC. 
The second constraint uses the optical properties of the two dust species to constrain the temperature ratio between carbon and silicate dust. When exposed to the general interstellar radiation field in the local solar neighborhood, carbon dust attains a temperature that is about 1.2 times higher than the silicate dust. 

Figure~\ref{sed_fits} shows the fits to the SEDs of the LMC and SMC. The best fit temperatures and masses are listed in Table \ref{params}. We derived a temperature of 22.4 K (26.9 K) for silicate (carbon) dust in the LMC, and 19.0 K (22.8 K) for silicate (carbon) dust in the SMC.The resulting total dust masses are $\rm 7.0\times10^{5}\: M_{\odot}$ and $\rm 2.0\times10^{5}\: M_{\odot}$ for the LMC and SMC, respectively. The individual masses for the silicate (carbon) dust components are $\rm 5.2\times10^{5}\: M_{\odot}$ ($\rm 1.8\times10^{5}\: M_{\odot}$) for the LMC, and $\rm 1.6\times10^{5}\: M_{\odot}$ ($\rm 4.0\times10^{4}\: M_{\odot}$) for the SMC. Our total mass estimates are consistent with the masses derived from a parametric fit to the SED of the LMC by \cite{gordon14} that yielded a total dust mass of $\sim 7.3\times 10^{5}$~\msun, somewhat lower that the values of $\sim 3.6\times 10^{6}$, $\sim 1.2\times 10^{6}$~\msun, and $\sim 1.7\times 10^{5}$~\msun, derived by \cite{bot10}, \cite{leroy07}, and \cite{bernard08}, respectively.

%==============================================================
\begin{figure*}
\epsscale{0.55} \plotone{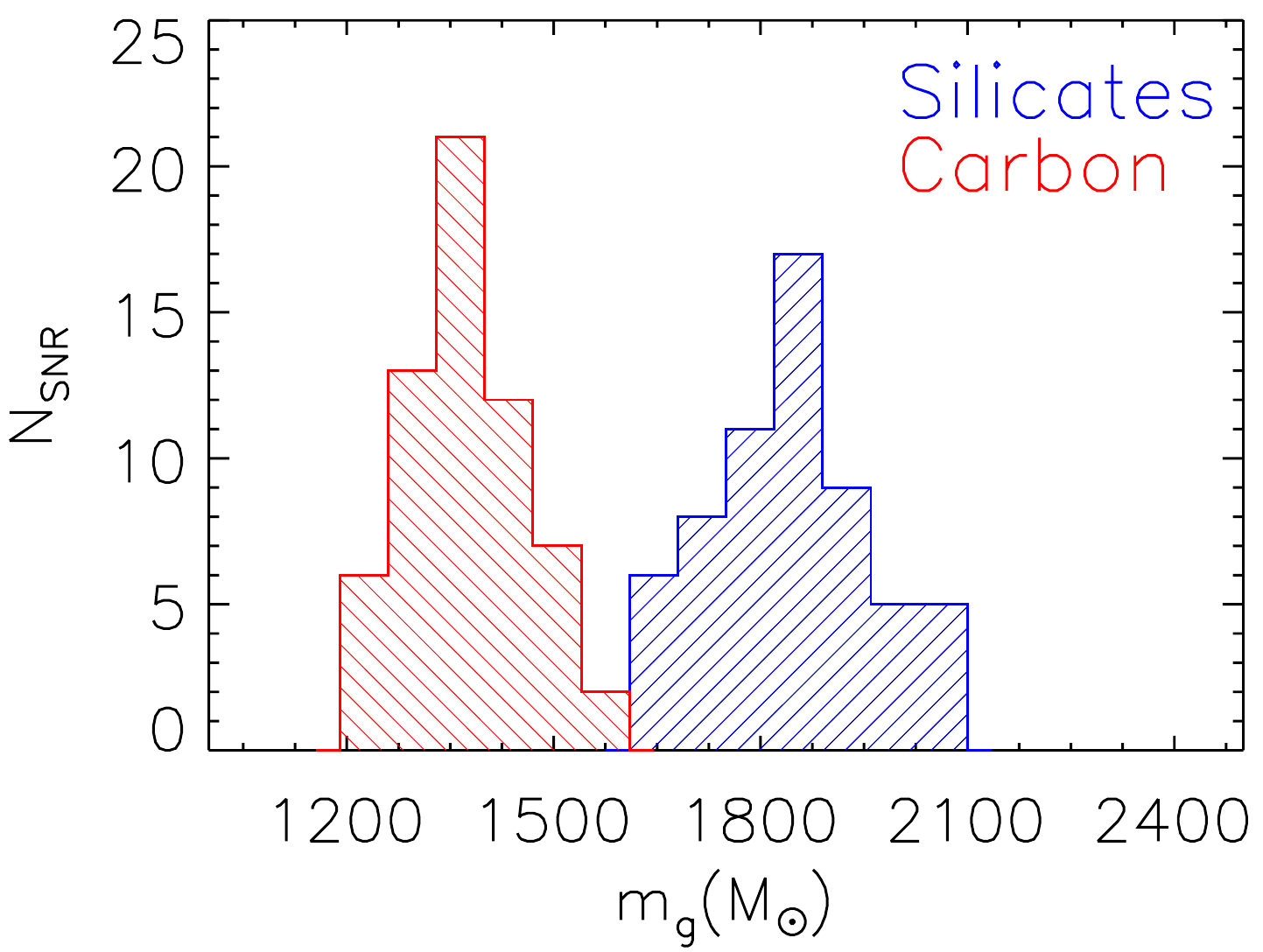}
\epsscale{0.55} \plotone{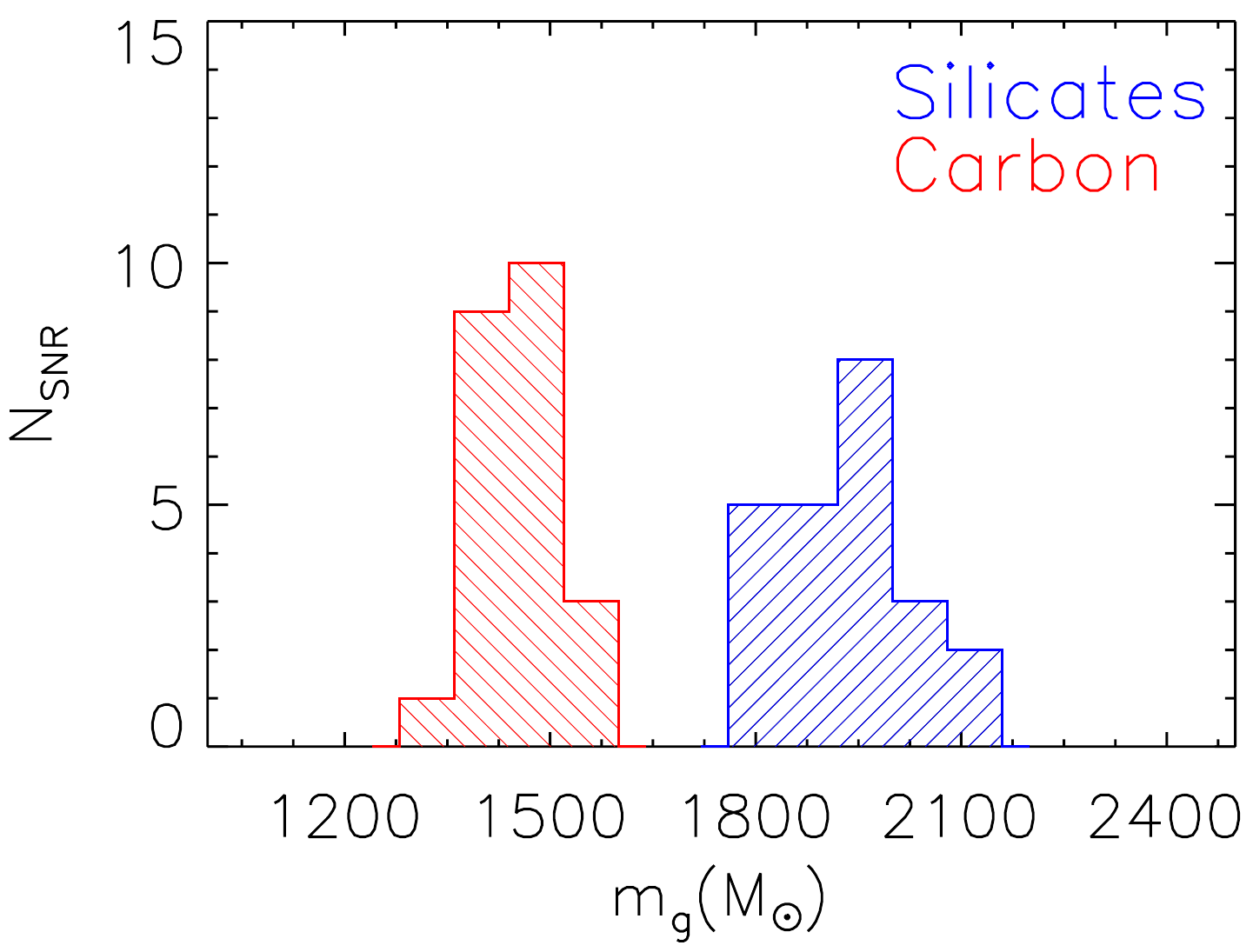}
\epsscale{0.55} \plotone{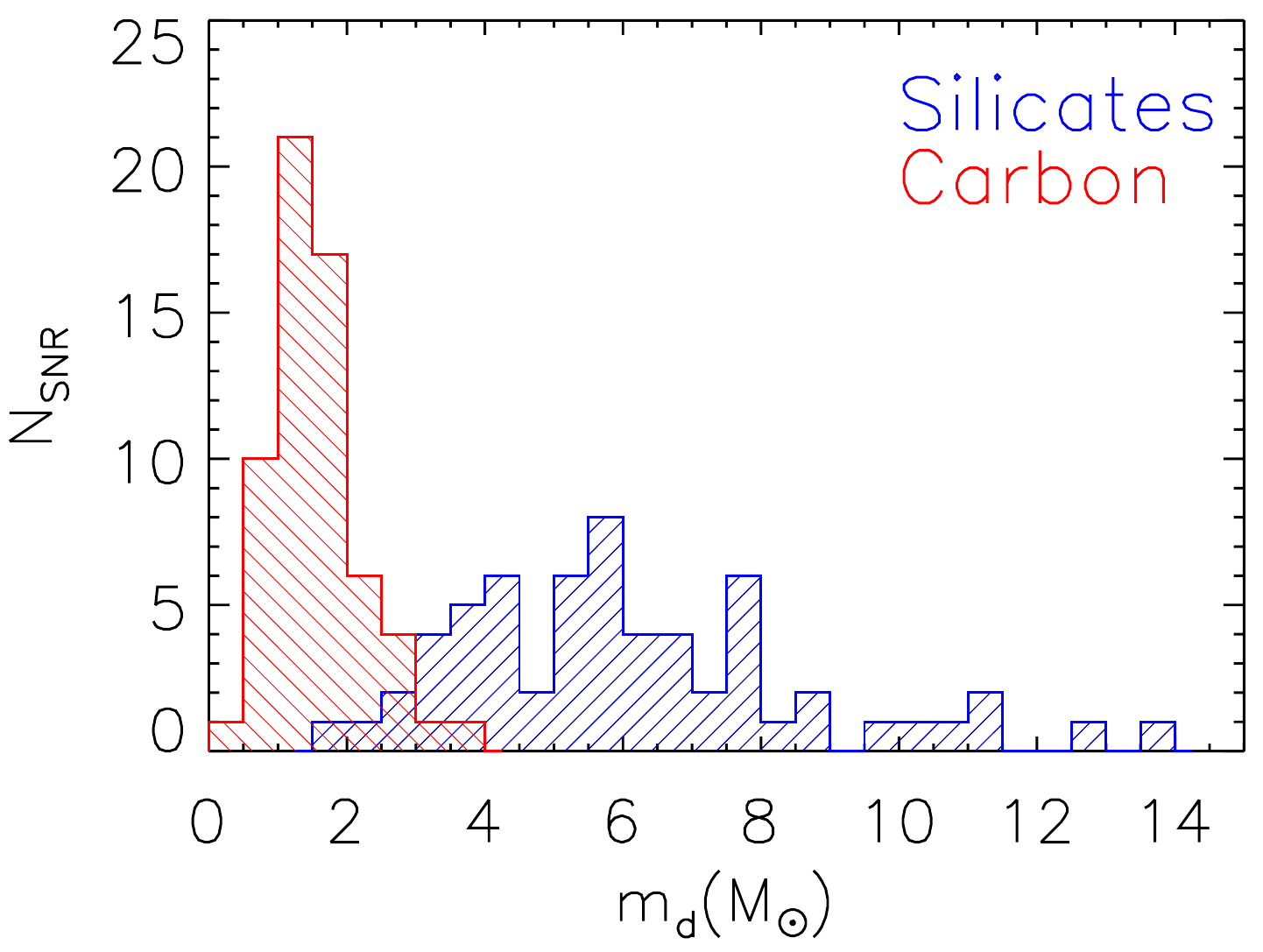}
\epsscale{0.55} \plotone{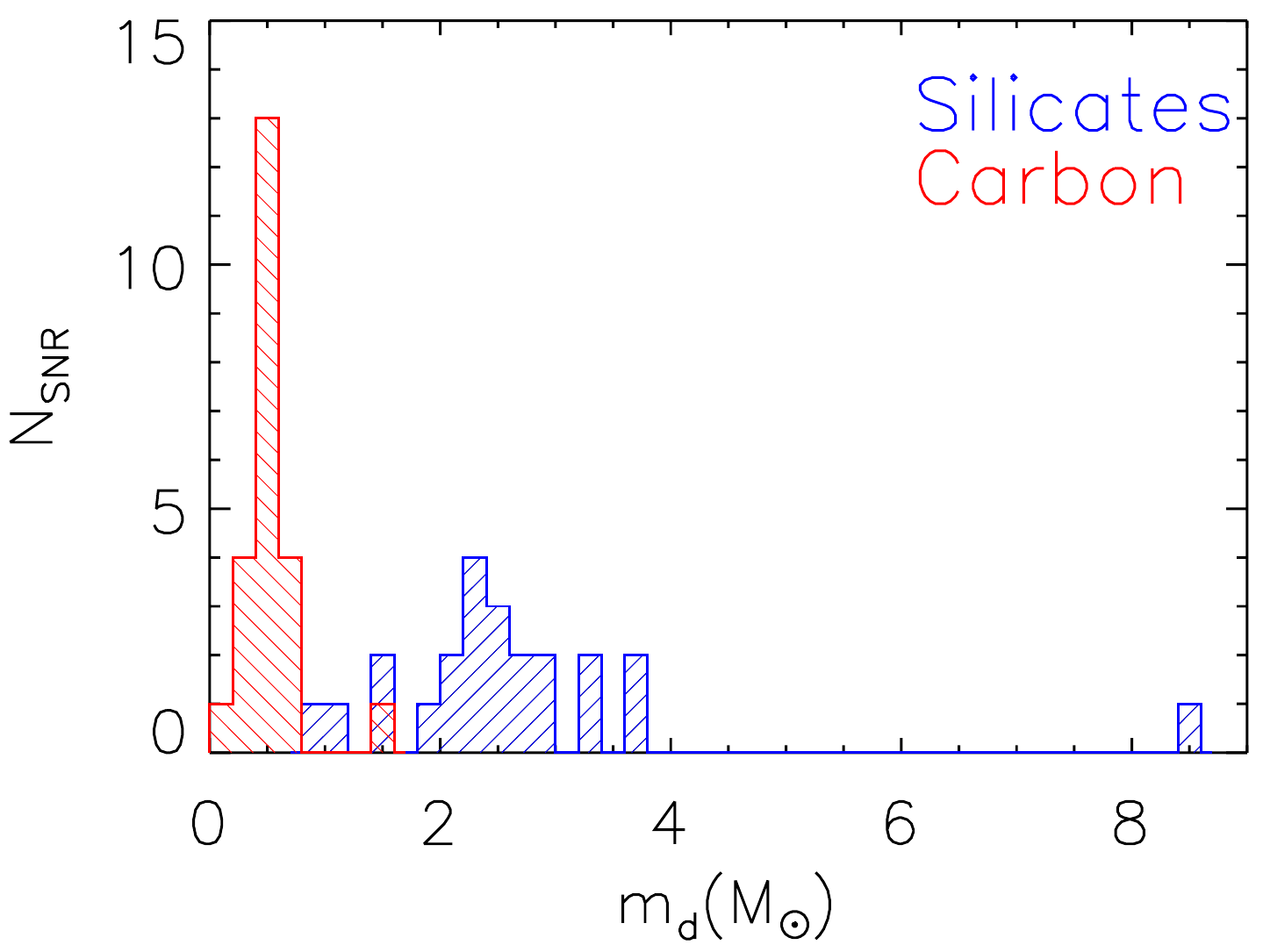}
\caption{\textbf{Top:} Histograms showing the effective gas mass \mg\ for the LMC (left) and SMC (right) SNRs using the silicate and carbon grain compositions. \textbf{Bottom:}  Histograms showing the total amount of dust mass destroyed $\mde$ by each SNR in the LMC (left) and SMC (right) for silicate and carbon grain compositions. A disk thickness of 0.4 kpc was assumed for the LMC, and 2.0 kpc for the SMC, and a metallicity factor $\zeta_m=0.3$. \label{panel1lmc}}
\vspace{0.1in}
\end{figure*}
%==============================================================

%--------------------------------------------
\subsection{Dust-to-Gas Mass Ratios Around the SNRs}
%-------------------------------------------- 
To calculate the D2G mass ratio in the ISM around each SNR, we extracted the 70--500 \micron\ infrared fluxes from the \textit{Herschel} PACS and SPIRE maps \citep{meixner13}, for the same annular regions surrounding each SNR as described in Section \ref{density}. We then fitted each of the 84 SEDs with a two-component dust model with the same carbon-to-silicate dust mass and temperature ratios used for the global SED fits (see Section \ref{tot_mass}). We divided the best-fit dust mass for each grain species individually by the total \small{HI} gas mass in each annular region. The resulting total D2G mass ratios are shown in Figure~\ref{nm_g2d} and listed in Tables~\ref{snrs_lmc} and \ref{snrs_smc}. Of the total dust mass, 74\% (26\%) is in the form of silicate (carbon) in the LMC, and 80\% (20\%) in the form of silicate (carbon) in the SMC. The D2G mass ratio does not depend on the choice of disk thickness, since the volume for the gas and dust masses simply cancels out. 

The determination of the dust mass in the HI gas assumes single temperatures for silicate and carbon dust, ignoring the possible presence of a colder dust component that may reside in molecular clouds. Allowing for the presence of a cold dust component may reduce the dust mass attributed to the HI gas. To check the effect of our approximation, we modeled the SED of two select regions characterized by the largest CO column density with two dust components. The temperature of the first, warm component was allowed to vary between 18 and 40~K, and that of the cold dust component between 6 and 18~K. 
The results showed that even when the cold component dominates the dust mass, it makes a negligible contribution to the total SED, and has therefore little effect on the dust mass attributed to the \small{HI} gas, which is the only relevant mass for calculating the grain destruction rate by SNRs. However, the presence of dust in molecular clouds can alter the total dust mass reservoir, $M_d$, and thus the dust lifetime, but since the molecular gas constitutes only 10\% of the gas mass in the MCs, the effect of cold dust on the dust lifetime is not significant.

Given the H~I mass of $4.0\times10^8$~\msun\ for the LMC and $2.5\times10^8$~\msun\ for the SMC \citep{roman-duval14}, and our  total \small{HI}-associated dust masses derived for the LMC and SMC, we find a global D2G ratios of $1.8\times10^{-3}$ and $8.0\times10^{-4}$ for the LMC and SMC, respectively. The average D2G ratios around the observed SNRs are a factor of $2-3$ higher,  $4.5\times10^{-3}$ and $1.7\times10^{-3}$ for the LMC and SMC, respectively. Table \ref{params} lists both the average global G2D value, and the average local value around the SNRs.

%--------------------------------------------
\subsection{Effective Swept-up Mass}
%--------------------------------------------

Given the model for the evolution of the SNR,  the grain destruction efficiency, and the ambient ISM density, we first calculated the value of $m_g$ for individual SNRs.
Figure~\ref{panel1lmc} shows the histograms of effective swept up gas mass $m_g$ for the entire SNR sample for both carbon and silicate grains, and an assumed gas disk thickness of 0.4 kpc for the LMC and 2.0 kpc for the SMC. The corresponding values for the individual SNRs are listed in Tables \ref{snrs_lmc} and \ref{snrs_smc}. The first important thing to note is that the distribution of $m_g$ is fairly narrow for any given disk thickness, implying that spatial variations in the \small{HI} column density in the MCs have a minor effect on $m_g$. The mean values and standard deviations of the distributions for different disc thickness values are listed in Table \ref{snrs_avg}. For any given disk thickness value, the standard deviation of $m_g$ is on the order of only 7\% of the mean. This essentially means that the dust mass destroyed by each SNR ($m_d$) varies only with the local D2G mass ratio, and that $m_d$ can roughly be estimated by taking the average value of $m_g$ in Table~\ref{snrs_avg}, and multiplying it by the local value of $Z_d$.
The dependence of the effective gas mass $m_g$ on density (i.e. disk thickness) is approximately a power-law of the form $m_g\propto n_0^{-0.107}$.

Since the SNR evolution depends on the metallicity of the surrounding ISM that affects the cooling of the gas, the effective swept-up gas mass $m_g$ also has a metallicity dependence. Based on \citet{cioffi88}, we find that $m_g\propto \zeta_m^{-0.15}$, where $\zeta_m$ is the metallicity normalized to its solar value, and taken to be equal to 0.3 for the MCs. For comparison, $m_g$ as a function of density for the Milky Way ($\zeta_m=1.0$) is shown in the right panel of Figure \ref{panel1lmc}. If the average metallicity around the observed SNRs in the MCs is higher than the average global metallicity, similar to the higher than average G2D, then the corresponding $m_g$ values will be lower by 10-13 \%.

%==============================================================
%--------------------------------------------
\subsection{Dust Mass Destroyed by Individual SNRs}
%--------------------------------------------
The total mass of dust destroyed by each SNR, $m_d$ is simply derived by multiplying $m_g$ by the total D2G mass ratio in the local SNR environment. The total mass of silicate and carbon dust destroyed by each SNR is then given by their relative contribution to the total dust mass.
The histogram of the resulting destroyed dust masses for each SNR for both carbon and silicate grains are shown in Figure \ref{panel1lmc} and listed in Tables \ref{snrs_lmc} and \ref{snrs_smc}. 
%==============================================================

The shapes of the distributions of the destroyed dust masses in Figure \ref{panel1lmc} are mainly determined by the spatial variations in the D2G mass ratio in the LMC and SMC. Since $m_g$ does not show a strong dependence on the gas density into which the SNR expands, the variations in $m_d$ are mainly caused by the variations in the D2G mass ratio.

\begin{figure*}
\epsscale{0.56} \plotone{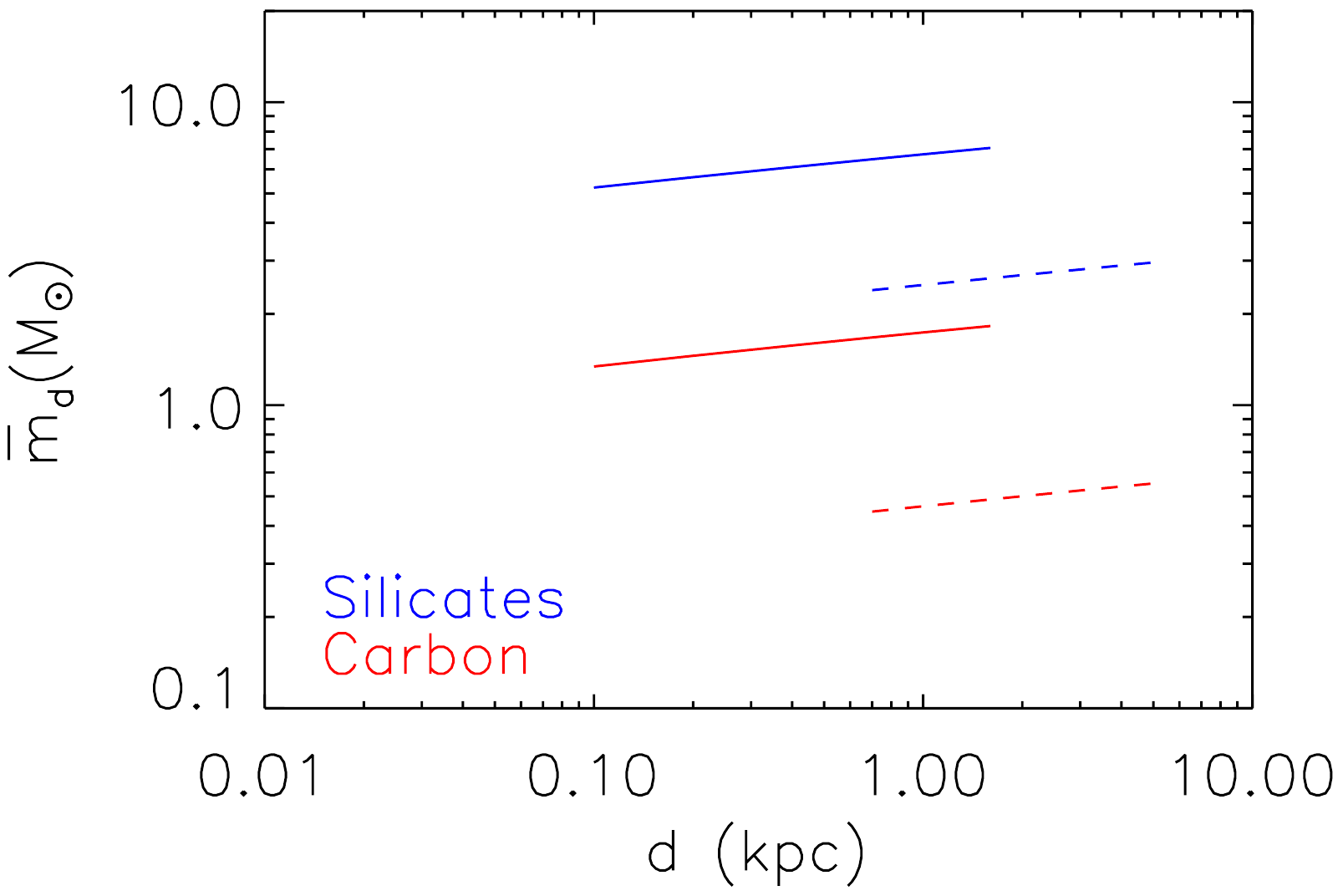}
\epsscale{0.55} \plotone{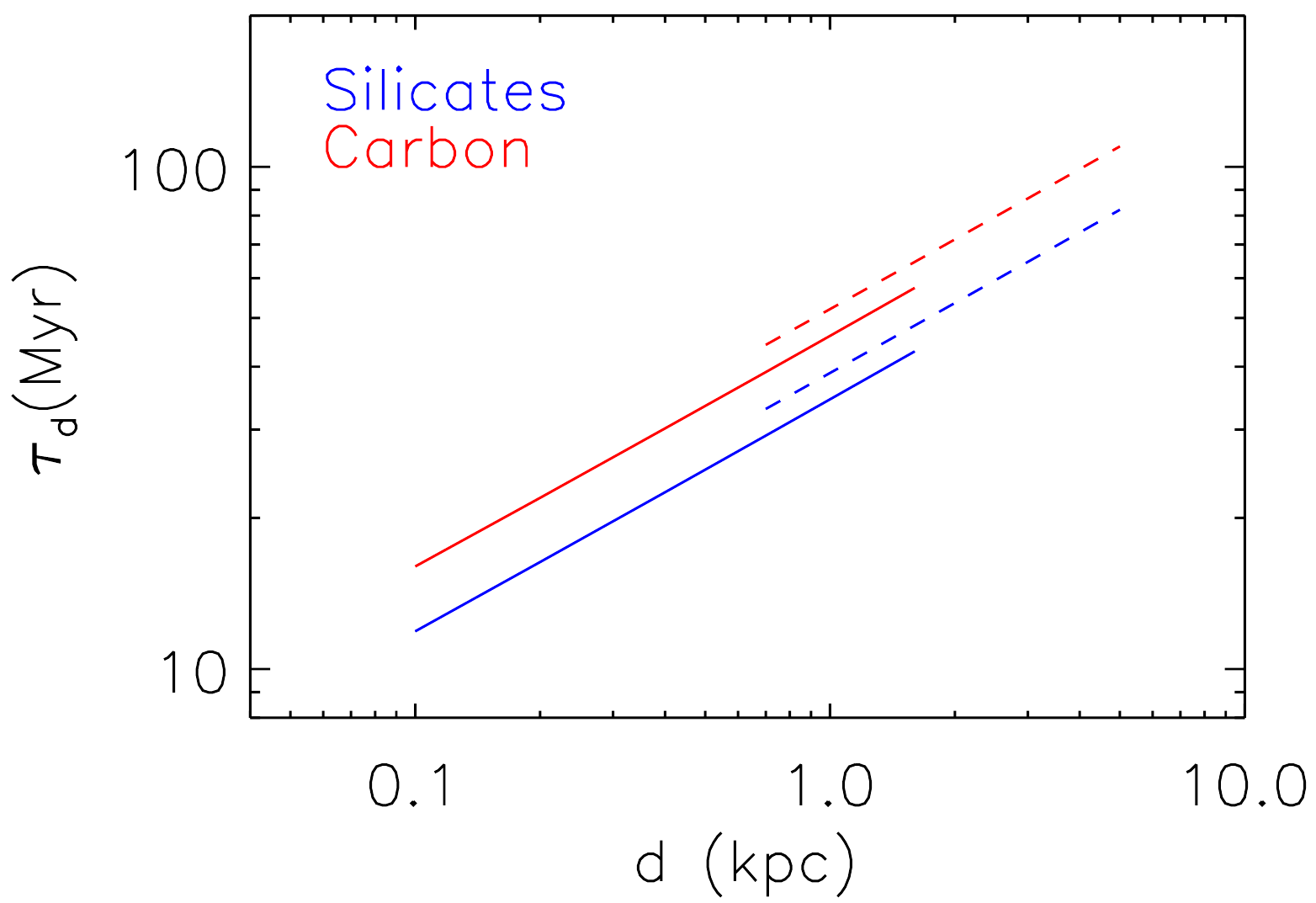}
\caption{Average mass of dust destroyed per SN (left) and the average dust lifetime (right), as a function of the disk thickness, assuming a metallicity factor $\zeta_m=0.3$. The red and blue lines represent silicate and carbon grains, respectively, while the solid (dashed) lines represent LMC (SMC) values. 
\label{panel2}}
\end{figure*}

The average destroyed dust mass values for all SNRs as a function of disk thickness (or density) are listed in Table \ref{snrs_avg} and shown in Figure \ref{panel2}.
The weak dependence of $\mde$ on gas density can be translated to a dependence on disk thickness $d$ as $\mde \propto d^{0.107}$. Since it is linearly dependent on $m_g$, it will have the same metallicity dependence, given by $\mde \propto \zeta_m^{-0.15}$.

%==============================================================
%==============================================================

%--------------------------------------------
\subsection{The Supernova Rate}
%--------------------------------------------

The supernova  rate can be derived from the observed number of SNRs, $N_{SNR}$, from the relation:
\begin{equation}
\label{nsnr}
R_{SN}={N_{SNR} \over \tau_{vis}},
\end{equation}
where $\tau_{vis}$ is the visibility time of the SNR. This rate is accounts for only the ``isolated" SNRs that we observe, and not any clustered SNRs that might have expanded inside giant or supergiant bubbles and escaped detection. As we will discuss in the next section, these clustered SNRs do not affect the dust destruction and production rates that we calculate, and are therefore not included in the SN rate.
\citet{maoz10} presented a physical model in which they adopted the epoch spent by a remnant in the Sedov (adiabatic) phase of its evolution as the remnant visibility time. We also adopt this definition of the visibility time for our analysis, but use the pressure-driven SNR model by \citet{cioffi88}.

\begin{figure}
\epsscale{1.2} \plotone{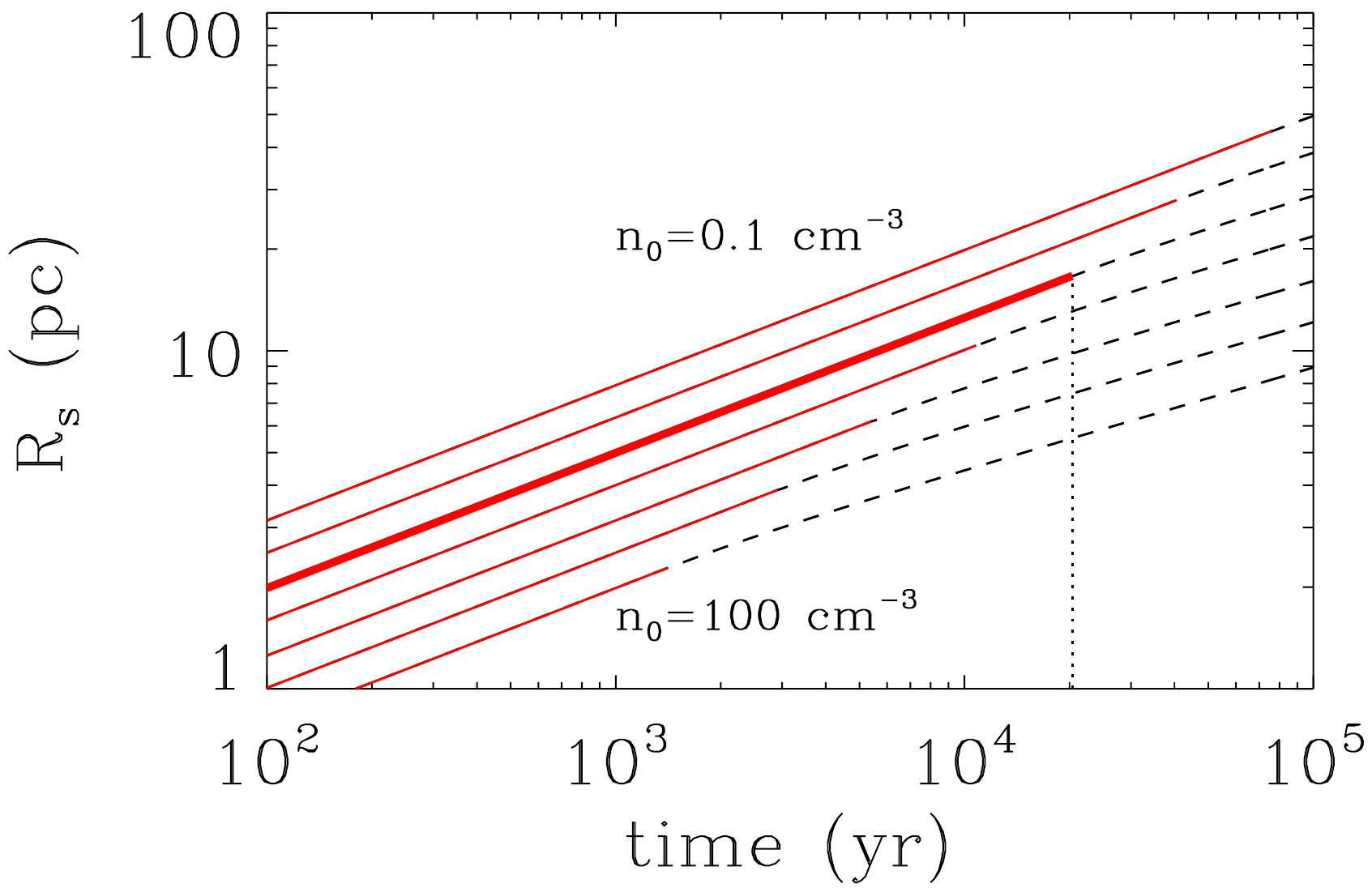}
%\epsscale{0.55} \plotone{mdest_snr_5.eps}
\caption{\label{snr_radius}Radius of a pressure driven SN as a function of time and ISM densities, for a metallicity of $\zeta_m=0.3$. The red and the black-dashed parts of the curve represent the Sedov and radiative phases of the remnant's evolution, respectively. The time spent in the Sedov phase is the visibility time of the remnant. The curves represent ISM densities of 0.1, 0.3, 1.0, 3.0, 10, 30, and 100~cm$^{-3}$. The bold line corresponds to a remnant expanding into a medium with $n_0 = 1$~cm$^{-3}$, which has a visibility time of $2\times 10^4$~yr.}
\end{figure}

Figure~\ref{snr_radius} shows the evolution of the SNR radius as a function of time for different ISM densities. The red segment of the curve represents the Sedov phase of the evolution of the remnant, and the dashed line the radiative phase.
The visibility time depends on the ISM density, and is given by: $\tau_{vis} = 2\times 10^4\, n_0^{-4/7}$~yr. 
For our adopted disk thickness values of 0.4 kpc for the LMC and 2.0 kpc for the SMC correspond to an average density around the SNRs of 2~\cc and 1~\cc for the LMC and SMC, respectively. This leads to an average visibility time of $16.3\times 10^3$~yr and $20.7\times 10^3$~yr for the LMC and SMC, respectively.  Our values are consistent with those of \citet{badenes10}, who derived a visibility time of $\approx (14-20)\times 10^3$~yr, based on various density tracers for the MCs,

Based on the above equation, $R_{SN}$ will have a $n_0^{-4/7}$ on density, and $d^{4/7}$ dependence on disk thickness. The values of $R_{SN}$ as a function of disk thickness for the LMC and SMC are listed in Table \ref{snrs_avg}. 
Given the number of confirmed SNRs, the average ISM densities for the nominal disk thickness values, and the visibility times, we derive SN rates of $\sim 3.75\times 10^{-3}$ and $\sim 1.11\times 10^{-3}$~yr$^{-1}$, for the LMC and SMC, respectively. These rates are lower than the \citet{harris09} SN rates, due to the fact that we only consider the observed isolated SNRs, and ignored cluttered SNR that give rise to giant and supergiant bubbles in the MCs.
The values of $R_{SN}$ as a function of disk thickness are listed in Table~\ref{snrs_avg}, and the nominal value and the functional dependence of $n_0$ (or $d$) are listed in Table \ref{params}.

%--------------------------------------------
\subsection{The Effect of SNR Clustering}
%--------------------------------------------
The presence of superbubbles in the MCs \citep{kim99,stanimirovic99} suggests that a fraction of all CCSNe created at any given time occur in a cluster environment. There are 103 giant and 23 supergiant shells catalogued in the LMC \citep{kim99}, and 495 giant and nine supergiant shells in the SMC \citep{stavelysmith97,stanimirovic99}. The presence of these bubbles suggests that there exists a population of CCSNe that has escaped detection. Consequently, any SFR derived from the rate of isolated SNRs will be lower than the absolute SFR, and for this reason, our derived SN rate for the observed SNRs is lower than the values found in \citet{harris09}.

Our rationale for using the lower SN rate, calculated from observed SNRs, is that clustered SNe have a negligible effect on the dust destruction and injection rates. Only the first massive star that explodes in the cluster will destroy the ambient dust. The remaining CCSNe will be expanding in a medium that has been cleared of dust by the first. Clustered CCSN have therefore a very low grain destruction rate compared to an identical number of isolated SNRs. The effect of correlated CCSNe therefore does not affect our dust destruction rate that was observationally derived from isolated SNRs. Similarly, clustered CCSNe do not significantly contribute to the dust injection rate, since any subsequent SN will destroy the dust produced by the previous one.

An important issue is if the observed isolated SNRs will overlap before their shock velocity drops below 50~km~s$^{-1}$, the threshold for grain destruction. If there is significant overlap in the volumes of the ISM that the SNRs sweep up, than we may have overestimated the dust destruction rate. In order to check this, we computed the SNR radii for a time at which all the dust destruction by the SNR has occurred. These radii are basically the radii at which each SNR has swept up an $m_g$ amount of gas. The resulting radii sizes range from $0.25-8$ times the current SNR radius, suggesting that some SNRs in the sample have already stopped destroying dust, while others will continue to destroy dust until they reach a radius several times larger than the current one. The extreme case is SN 1987A, which will continue to destroy the ambient dust up to a radius of $\sim80$ times the current SNR radius. We overlaid these evolved SNRs sizes onto the H~I maps of the LMC and SMC, and found that there is no overlap in the encompassing volumes, and that the dust destruction rate has not been overestimated due to this effect.

%--------------------------------------------
\section{Dust Lifetime and Destruction rates in the Magellanic Clouds}\label{lifetimes}
%--------------------------------------------
The dust lifetime, $\tau_d$ (Equation \ref{taud}), calculated using the total dust mass for each grain species, the supernova rate $\rm R_{SN}$, and the average destroyed dust mass per SN ($\mde$), are all summarized in Table \ref{params}. For the chosen gas disk thickness values and a metallicity of $\zeta_m=0.3$, the dust lifetimes for the LMC and SMC are $22 \pm 13$ Myr ($30 \pm 17$ Myr) and $\rm 54 \pm 32$~Myr ($\rm 54 \pm 32$~Myr) for silicate (carbon dust), respectively. 
This corresponding dust destruction rates are $2.3\times10^{-2}$~\myr, ($5.9\times10^{-3}$~\myr) for silicate (carbon) dust in the LMC, and $3.0\times10^{-3}$~\myr\  ($5.6\times10^{-4}$~\myr) for silicate (carbon) dust in the SMC.
The dependence of the dust lifetime and destruction rate on disk thickness  and metallicity factor is $\rm d^{0.464}\zeta_m^{-0.207}$ and $\rm d^{-0.464}\zeta_m^{0.207}$, respectively (see Tables~\ref{snrs_avg} and Table~\ref{params}). The plot of $\tau_d$ as a function of $d$ for the entire range of disk thickness values is shown in Figure~\ref{panel2}. The dependence of the dust lifetimes and destruction rates on metallicity is also listed in Table~\ref{params}.

%====================================
\subsection{Comparison to the Milky Way}
%====================================
In a uniform ISM, with a constant D2G mass ratio, the dust lifetime is independent of the total dust mass, and can be written as: $\tau_d = M_g/(\mge\, R_{SN})$. Since \mg\ is only a weak function of the ambient density in which an SNR is expanding, and the LMC values of $M_g$ and $R_{SN}$ are similar to that in the solar neighborhood, one would expect similar dust lifetimes in these two systems. However, our results show that the lifetimes of the silicate and carbon dust in the LMC and SMC are about an order of magnitude less than the $\sim 400$ and $\sim 200$~Myr derived by \cite{jones11} for the respective dust species in the solar neighborhood.
As shown below, the reasons for this difference is the lower dust mass in the MCs, that the D2G mass ratio is not uniform, and that SNRs expand preferentially into an ISM with a higher than average D2G mass ratio.

The gas surface density and total (Type~Ia and CCSN) rates in the solar neighborhood are $\sim10$~\msun\ pc$^{-2}$ and $\sim 0.016$~pc$^{-2}$~Gyr$^{-1}$, respectively \citep{rana91,dickey93,cappellaro96,tammann94}, giving an $M_g/R_{SN}$ ratio of $6.2\times10^{11}$~\msun~yr.  
For an ISM density of 1~cm$^{-3}$, the value used by \cite{jones11} to calculate the dust lifetimes, and a metallicity $\zeta_m=1.0$, we get values of \mg=1600~\msun\ and 1200~\msun, for silicates and amorphous carbon, respectively.  Using eq.~(\ref{taud}) we derive dust lifetimes of 375~Myr and 500~Myr for these respective dust species. Our global approach is therefore capable of reproducing the lifetimes derived by the more detailed models to within a factor of two.   

For the LMC and SMC, the $M_g/R_{SN}$ ratio is equal to $1.1\times10^{11}$~\msun~yr and  $2.3\times10^{11}$~\msun~yr, respectively, where we adopted an H~I mass of $4.0\times10^8$~\msun\ for the LMC and $2.5\times10^8$~\msun\ for the SMC \citep{roman-duval14}. Even in a uniform ISM with a constant D2G mass ratio, we would expect LMC and SMC dust lifetimes to be about $\sim6$ and $\sim3$ times lower than those of the solar neighborhood. 

An additional difference between the dust lifetimes in the MCs and solar neighborhood stems from the assumption that the D2G mass ratio is constant throughout the ISM.
The average D2G mass ratio, $M_d/M_g$ is equal to $1.8\times 10^{-3}$ and $8.0\times 10^{-4}$ for the LMC and SMC, respectively. Table~\ref{params} shows that the D2G mass ratios of the ISM into which the SNRs are expanding are larger by an average factor of $2-3$ compared to the global D2G mass ratio, even though the standard deviation on these values indicates a significant spread (see Table~\ref{params}). Since core-collapse SN progenitors explode in the vicinity of other massive stars, it perhaps may not be surprising for their SNRs to be found in ISM environments that are more enriched in metals and dust. Since the average SN rate in the MCs is comparable to that in the solar neighborhood, we  conclude that the lower dust lifetimes in the MCs are the results of their lower total dust mass and the fact that  a significant fraction of SNRs seems to expand into an ISM with a higher than average D2G mass ratio. 

  %==============================================================

%==============================================================

%====================================
\section{APPLICATION TO OTHER GALAXIES AND \\dust evolution models}\label{application}
%====================================

The SN rate that we are using to calculate dust destruction lifetimes, listed in Tables~\ref{snrs_avg} and \ref{params}, is the SN rate derived from the isolated SNRs that we are currently observing, including both CC and Type Ia explosions. This SN rate is lower than the absolute SN rate that one may derive from the SFR of \citet{harris09}, due to the fact that we leave out any correlated SNe that may have occurred inside large bubbles, and are not expected to destroy a significant amount of dust. 

In order to apply our results for another epoch in the MCs or another galaxy for which the isolated SNR population is not resolved, one needs to first calculate the nominal CCSN rate from the SFR given by: $R_{SN}(t)=\psi(t)/m_{\star}$, where $\psi(t)$ is the SFR, and $m_{\star}$ is the mass of stars generated per CCSN event \citep{dwek11}. To calculate the effective rate of dust destroying SNR we need to take the clustering of CCSN into account, since only field SNRs will destroy the dust, and the presence of Type~Ia SNR. The rate of SN that destroy the dust at any epoch $t$, is then given by:
\begin{equation}
\label{ }
R_{SN}(t) = \left[{\psi(t) \over m_{\star}}\right]\, (1-f_{cl}+f_{Ia})
\end{equation}
 where $f_{cl}$ is the fraction of CCSN that are clustered, and  $f_{Ia}$ is the fraction of Type~Ia to CCSN.

Given a SN rate, Equations~(\ref{taud}) and (\ref{dmd_dt}) can be used to calculate the dust lifetime and destruction rate, where the mass destroyed by an average SN can be estimated by $\mde = \overline{Z_d}\, \mge$. Here,   $\overline{Z_d}$ is the average D2G mass ratio of the ISM into which the SNe expand, which can be higher than the global D2G mass ratio of the galaxy.

Chemical evolution models usually take \mg\ to be constant because of its weak dependence of ambient density. However \mg\ also depends on the metallicity of the ambient medium into which the SNRs expand, since it determines the postshock cooling of the gas, and therefore the evolution of the remnant. These two effects should be considered in future dust evolution models.
The value of $\mge$ should be adjusted for the desired metallicity and average density of the galaxy, according to Table~\ref{params}.

%============================================
\section{DUST PRODUCTION RATES BY \\CCSNe AND AGB STARS} \label{production}
%============================================
%--------------------------------------------
\subsection{Dust Production Rate by Type~Ia SN}
%--------------------------------------------

Theoretically, calculations by \cite{nozawa11} show that dust can form in these objects. However, the results show that because of the high expansion velocity and low mass of the ejecta, the resulting grain sizes are very small ($\lesssim 100$ \AA). Furthermore, the ejecta is less likely to be clumpy, so that all dust that may have formed is expected to be destroyed by the reverse shock \cite{nozawa11}.

Observationally, searches for dust in remnants of Type~Ia SNe, including  Kepler \citep{blair07, williams12}, RCW 86 \citep{williams11a}, SN 1006 \citep{winkler13}, and Tycho \citep{williams13}, have yielded negative results. That no significant grain formation takes place in Type Ia SNe was later confirmed by {\it Herschel} observations of Kepler and Tycho \citep{gomez12b}. We therefore ignore Type~Ia SN as sources of dust.

%--------------------------------------------
\subsection{Dust Production Rate by CCSNe}
%--------------------------------------------
In contrast to Type~Ia SNe, there is considerable evidence for the formation of dust in the ejecta of CCSNe. CCSNe have relatively short main sequence lifetimes, $< 40$~Myr for a 8~\msun\ progenitor, compared to the lifetime of an average AGB star. We will therefore assume that the CCSN event occurs promptly after the birth of its progenitor. The dust production rate can then be written as:
\begin{equation}
\label{ }
\left[{dM_d\over dt}\right]_{CCSN} = {\overline Y_d}\, R_{CCSN}
\end{equation}
where ${\overline Y_d}$ is the CCSN yield averaged over the stellar initial mass function (IMF), and $R_{CCSN} = N_{CCSN}/\tau_{vis}$ is the rate of CCSNe. Deriving the value of  $R_{CCSN}$ therefore requires the subtraction of SNR that are the result of Type~Ia events from the SNR sample. 
\cite{maoz10} estimated the fraction, $f_{Ia}$, of Type~Ia SNR in the sample to be between 0.1 and 0.5. To be definitive, we will adopt an average value of $f_{Ia}=0.30$ in all our calculations.

Determining the dust yield from CCSNe is complicated by the fact that observations taken shortly after the explosion usually sample only the hot dust, and may therefore be missing any cold dust component that may be hidden in optically thick clumps.
Detection of dust during the remnant phase is necessarily limited to young remnants, before their ejecta has mixed with the ISM. 

\spitzer\ and \textit{Herschel} observations of SNRs opened new spectral windows that cover a wide range of dust temperatures and dust emission features, enabling determination of dust composition and heating mechanisms. Surveys of young, unmixed remnants with these satellites revealed $\sim 0.01-0.2$~\msun\ of dust in the ejecta of SNRs such as the Crab Nebula, Cas~A, G292+1.8, E0102, and G11.2-0.3 \citep{koo07,rho08,rho09b,sandstrom09,barlow10,ghavamian12,gomez12a,temim13,arendt14}. 

The largest mass of SN condensed dust was found in SN1987A. \spitzer\ and \textit{Herschel} observations revealed $\sim 0.5-0.7$~\msun\ of dust \citep{matsuura11}, which was subsequently spatially resolved with ALMA and definitively associated with the expanding SN ejecta \citep{indebetouw14, zanardo14}. The large mass of dust found in SN1987A suggests that almost all of the refractory elements precipitated out of the gas and formed dust with nearly 100\% efficiency.

%============================================
\begin{figure}
\epsscale{1.0} \plotone{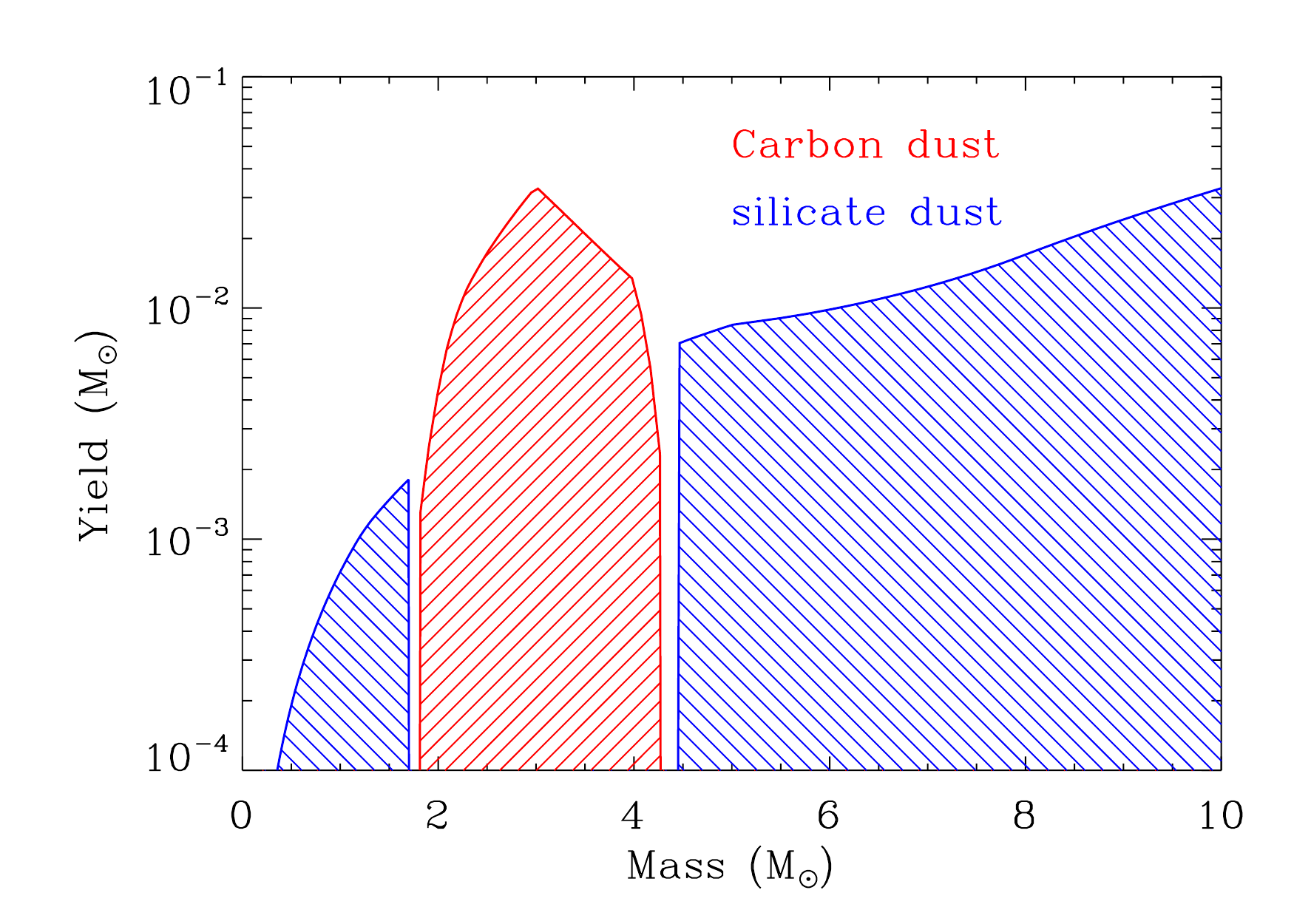}
%\epsscale{0.55} \plotone{yield_mg.pdf}
\epsscale{1.0} \plotone{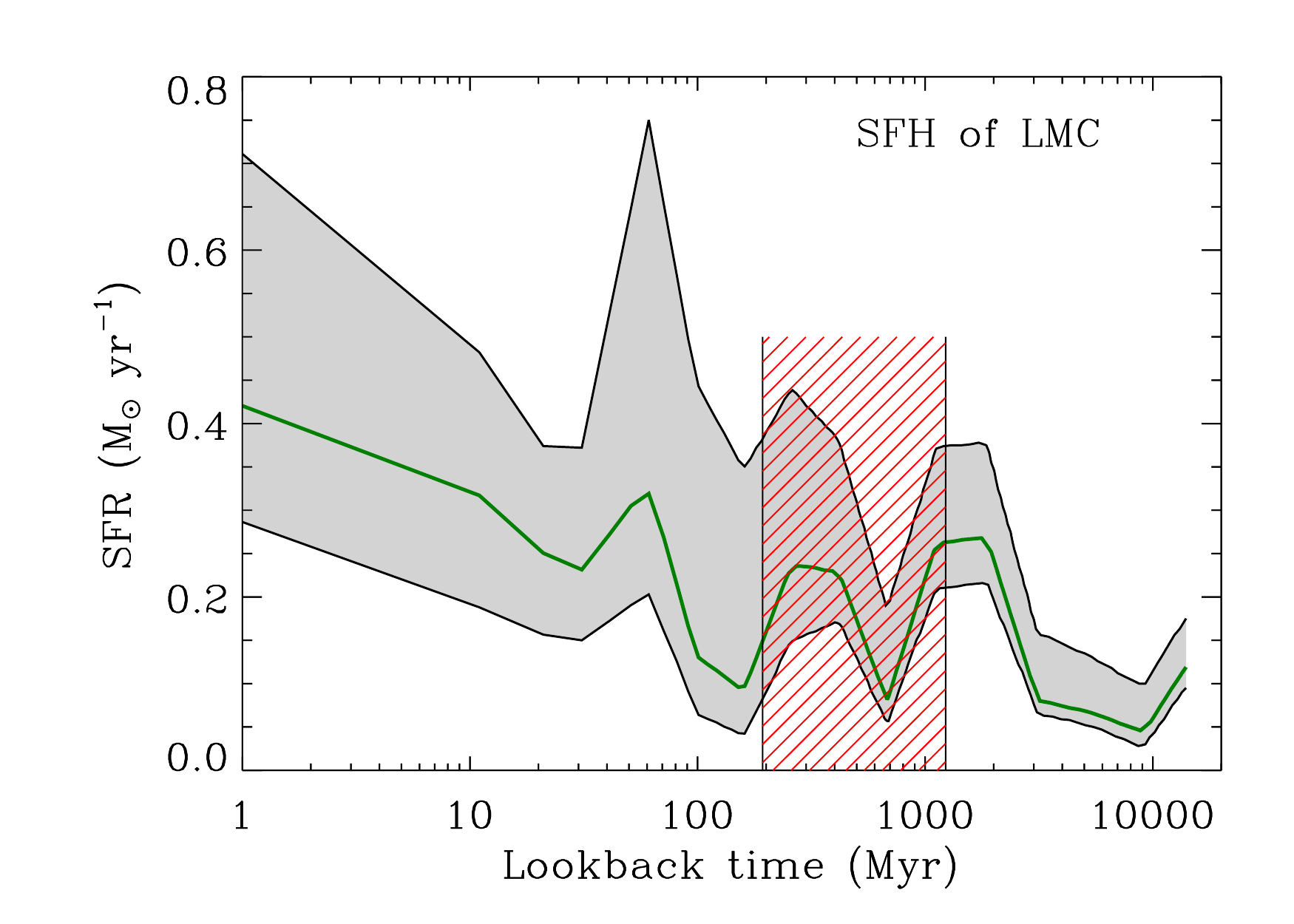}
\epsscale{1.0} \plotone{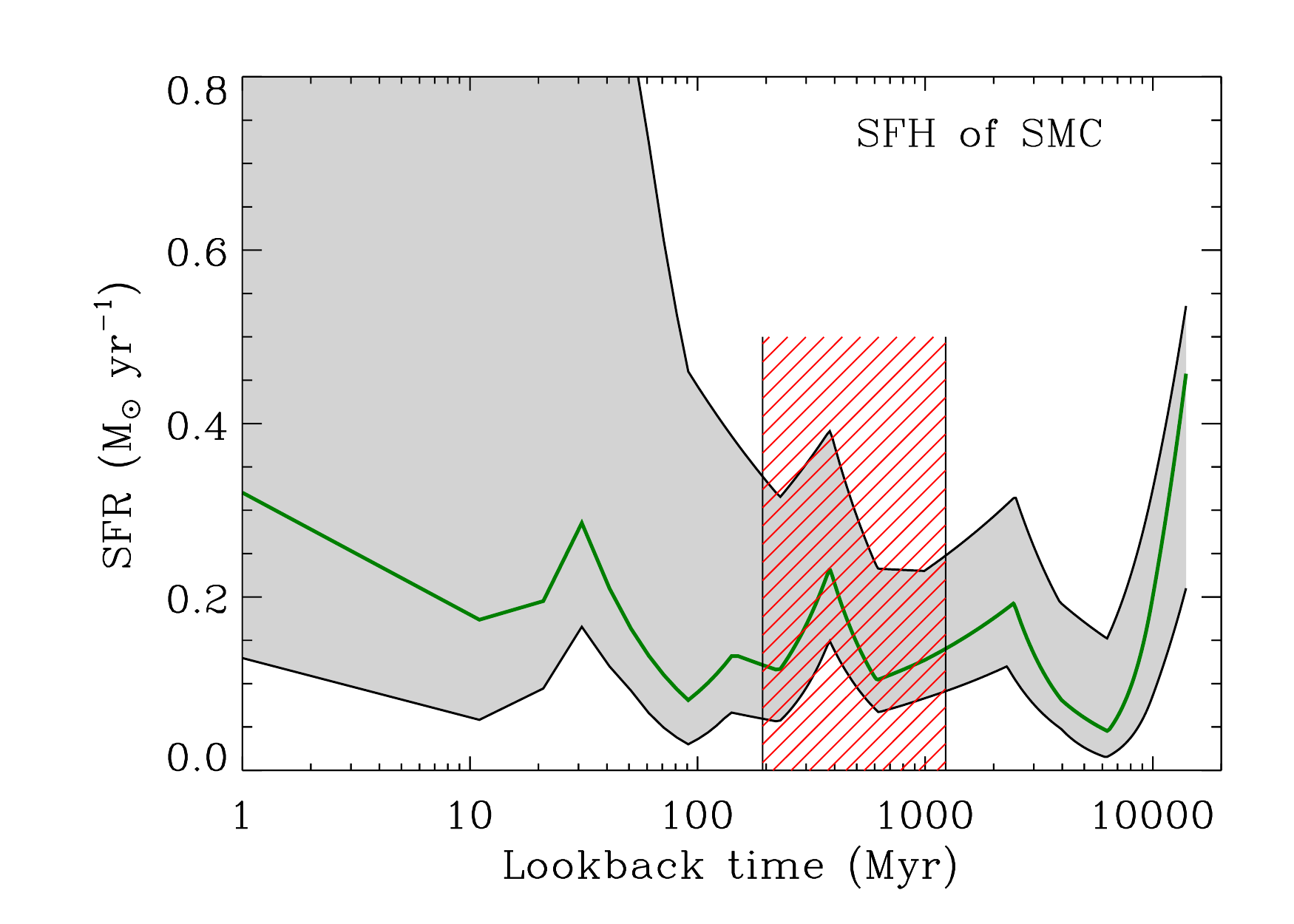}
\caption{\label{yield} \textbf{Top:} The dust yield in AGB stars using the \citet{karakas10} yields. The yields assume a 100\% condensation efficiency in the sources, and that all the carbon (oxygen) is locked up in dust when the C/O ratio is $>$ 1 ($<$ 1). 
\textbf{Middle and bottom:} The star formation histories of the Large and Small Magellanic Clouds after \cite{harris09}.
The shaded area depicts the birth time of carbon-rich AGB stars that are forming carbon dust at the present epoch.}
\end{figure}
%============================================

The yields presented above suggest that only a fraction of the condensible elements in the ejecta of CCSNe form dust. The separate yields of carbon and silicate dust are only known, with great uncertainty, for select remnants. The uncertainty in the dust composition stems from the fact that most of the dust is cold and emits at far-IR wavelengths where there are no distinguishing solid state features.  Using all the information above, we adopt a CCSN dust formation efficiency of 20\%, and use the theoretically derived elemental yields \cite{woosley95} to calculate the dust yield in massive stars. These yields are tabulated for 100\% condensation efficiencies in Table 2 of \cite{dwek07b}.  We average the yields over a mass function of the form $\rm M_{*}^{-2.35}$, for stars between 8--40 $\rm M_{\odot}$. The final IMF-averaged dust yields that we calculate for a 100\% condensation efficiency for silicate and carbon grains are 0.5 $\rm M_{\odot}$ and 0.15 $\rm M_{\odot}$, respectively. 

With $R_{CCSN}=f_{Ia}\, N_{SNR}/tau_{vis}$ and the visibility times listed in Table~\ref{params}, we get a CCSN rate of $2.6\times10^{-3}$~yr$^{-1}$ and $7.8\times10^{-4}$~yr$^{-1}$ for the LMC and SMC, respectively. 
The resulting dust injection rates are $1.3\times10^{-3}$~\msun/yr and $3.9\times10^{-4}$~\msun/yr for silicates and carbon grains in the LMC, and $3.9\times10^{-4}$~\msun/yr  and $1.2\times10^{-4}$~\msun/yr for silicate and carbon grains in the SMC. The calculated dust injection rates are listed in Table~\ref{rates}.

%---------------------------------------
\subsection{AGB Dust Injection Rates}
%---------------------------------------
Low mass stars ($M \leq 8$~\msun) form dust in their winds during the aymptotic giant branch (AGB) phase of their evolution.  The dust is injected into the ISM after the stars evolve off the main sequence (MS). The delayed dust injection rate by AGB stars, $[dM_d(t)/ dt]_{agb}$ is given by: \citet[e.g.][]{dwek98, dwek11a, zhukovska08a, calura10}:
\begin{equation}
\label{ }
\left[{dM_d(t)\over dt}\right]_{agb} = \int_{m_1}^{m_w}\ Y_{agb}(m)\, {\psi[t-\tau_{MS}(m)]\over \left<m\right>}\, \phi(m)\, dm
\end{equation}
where $\tau_{MS}(m)$ is the MS lifetime of a star of mass $m$, $\left<m\right>$ is the IMF-averaged stellar mass, $m_1$ the mass of the lowest mass star that evolved off the MS by time $t$, $Y_{agb}(m)$ is the total dust yield in AGB stars, and $m_w$ is the upper mass limit of AGB stars, that is, the lower mass limit of stars that become CCSNe. 

Figure~\ref{yield} shows the carbon and silicate dust yields in AGB stars versus stellar mass. Elemental yields were taken from \cite{karakas07}, \citet{nanni13}, and Marigo (private communication), for a metallicity of Z = 0.008. Stars with C/O ratios $>1$ were assumed to produce only carbon dust, and stars with C/O ratios $<1$ were assumed to produce only silicate dust. The yields presented in the figure also assume a 100\% efficiency in the condensation process. The prescription for calculating the dust yields were presented in \cite{dwek98}. More realistic models, and a comprehensive comparison of the different AGB yields are presented in \cite{schneider14}. All models agree that at the LMC metallicity, carbon dust is produced by stars in the $\sim 1.5-3.5$~\msun\ mass range, with an average yield of $\sim(1-10)\times 10^{-3}$~\msun. 

Because of the delay time between the stellar birth and the epoch of dust injection, the dust production rate by AGB stars requires knowledge of the star formation history of the MCs. For sake of consistency with the calculated dust production rate by CCSNe, we will first calculate the current SFR in the MCs from the CCSN rate.
The relation between the two is given by:
\begin{equation}
\label{ }
\psi = m_{\star}\, R_{CCSN}
\end{equation} 
Assuming that all stars with masses above 8~\msun\ end up as CCSNe (e.g. Heger et al. 2003), the value of $m_{\star}$ depends on the stellar IMF, and is equal to 135~\msun\ for a Salpeter, and 90~\msun\ for a Kroupa IMF \citep{dwek11a}.
For a Salpeter IMF we get that the SFR is 0.35 \myr\ and 0.1 \myr\ for the LMC and SMC respectively, which is consistent with the values derived by \cite{harris09}. 

We will therefore adopt their nominal SFH for calculating dust injection rate by AGB stars in the MCs.
The middle and bottom panels of Figure~\ref{yield} show the star formation rate of the LMC and SMC versus lookback time \citep{harris09}. The shaded area depicts the epoch during which carbon stars in the 1.5--3.5~\msun\ mass range contributed to the currently observed injection rate of carbon dust.

Table~\ref{rates} compares our calculated injection rates of carbon and silicate dust in the LMC and SMC to observations. 
Not surprisingly, the calculated rates for the LMC are higher by a factor of $\sim 5$ from the observed range of values, since we adopted a condensation efficiency of 100\%. The observations therefore suggest that only $\sim 20$\% of the refractory elements condense in the ejecta. The discrepancy for the SMC is much higher, suggesting an unusually low condensation efficiency.

\section{ASTROPHYSICAL IMPLICATIONS} \label{implications}

Table \ref{rates} compares the carbon and silicate dust production and destruction rates derived in this paper. The results show that in both the LMC and SMC, the rate of grain destruction by SNRs greatly exceeds the rate of dust injection by AGB stars and CCSNe. This imbalance stems fundamentally from the fact that CCSNe destroy more dust during the remnant phase of their evolution than they produce in their ejecta shortly after core collapse.
Only in the very early universe ($z \gtrsim 9$), when the ambient dust-to-gas mass ratio was very low, are SN net producers of interstellar dust \citep{dwek14}
This imbalance between the formation and production rates of dust is not limited to the MCs \citep[see also][]{zhukovska13, schneider14}. It also exists in the local solar neighborhood, in galaxies in the local universe \citep{dwek98, zhukovska08a, calura10, galliano08a}, and in the high-redshift universe \citep{dwek07b, valiante11, dwek07b, dwek11a, dwek11b, gall11a, gall11b, michalowski11}.

One possibility for the discrepancy may be due to a significant overestimate of the grain destruction efficiency in the ISM. For example,  \citet{jones96} find that increasing the density $n_0$ from 0.25 $\rm cm^{-3}$ to 25 $\rm cm^{-3}$ increases the dust destruction efficiency by up to a factor of two for a shock velocity of 100 $\rm km\: s^{-1}$. However, integration over all shock velocities leads to a net increase in the dust destruction rate of only a few percent (Slavin, private communication, in preparation). A larger uncertainty is likely caused by the assumed initial grain size distribution, since any weighting towards the smaller grains would lead to a higher dust destruction efficiency.
Another reason for an overestimate of the dust lifetime may be due to our assumption that the SNRs expand into a homogeneous ISM. SNR expanding in a three-phase ISM, dominated by a low-density hot gas will leave most of the dust residing in the dense phase of the ISM intact \citep{dwek79, dwek07b}, but since molecular gas constitutes only 10\% of the gas mass in the MCs, this effect should not significantly increasing the dust lifetime. More detailed models following the destruction of dust in a clumpy medium are currently being performed by \cite{slavin14}.  

Alternatively, most dust giving rise to the observed IR emission in these galaxies might have grown by accretion onto surviving thermally-condensed cores in the dense ISM. This possibility is supported by the positive correlation between an element's condensation temperature and interstellar depletion \citep{field74}, that may also be interpreted as a trend reflecting the accretion efficiency in molecular clouds \citep{snow75}. 
The composition of interstellar dust must therefore reflect that of composite interstellar grains,  composed of refractory silicate or carbon cores and accreted refractory organic material \citep{greenberg95, li97, zubko04, jones13}.
Models consisting of composite dust have been successful in reproducing the observed interstellar extinction, diffuse IR emission, and interstellar abundance constraints in the local solar neighborhood \citep{zubko04}. Such  dust model will require the reevaluation of dust destruction rates and lifetimes, and the inclusion of accretion as an additional source of dust.

\section{SUMMARY}

We calculated the rate of grain destruction by SNRs in the LMC and SMC by modeling the evolution for a nearly complete sample of SNRs, using observationally determined values for the gas density and dust content around each SNR.
We find that the average dust mass destroyed by an SNR is $6.1\pm2.6$~\msun\ ($1.6\pm0.7$) of silicate (carbon) dust in the LMC, and $2.7\pm1.5$~\msun\ ($0.6\pm0.3$~\msun) of silicate (carbon) dust in the SMC. The quoted  values assume a disk thickness of 0.4 kpc for the LMC, and 2.0 kpc for the SMC, and an average metallicity factor for the MCs of $\zeta_m=0.3$.
The derived dust lifetimes are $22\pm13$ Myr ($30\pm17$ Myr) for silicate (carbon) grains in the LMC, and $54\pm32$ Myr ($72\pm43$ Myr) for silicate (carbon) grains in the SMC.
These values correspond to dust destruction rates of $(2.3\pm1.3)\times10^{-2}$~\myr\ ($(5.9\pm3.4)\times10^{-3}$~\myr) for silicate (carbon) grains in the LMC, and $(3.0\pm1.8)\times10^{-3}$~\myr\ ($(5.6\pm3.3)\times10^{-4}$~\myr) for silicate (carbon) grains in the SMC. 
The dust lifetimes and dust destruction rates in the MCs have a $n_0^{-0.464}\zeta_m^{-0.207}$ and $n_0^{0.464}\zeta_m^{0.207}$ dependence on the gas density and metallicity of the ISM into which the SNRs expand, respectively.

We also show that in general, the effective swept-up gas mass $\mge$ has a $n_0^{-0.107}\zeta_m^{-0.15}$ dependence on gas density and metallicity. The dependence of the dust lifetime and destruction rate on $\mge$ and these parameters should be taken into account in any future dust evolution models.

The dust lifetimes for the MCs are several times lower than those for the Milky Way, which can be explained by the combined effect of the lower total dust mass in the MCs, but also the fact that the isolated SNRs that have the most impact on dust destruction occur in regions with higher than average D2G mass ratios. 
We find that the derived dust destruction rates are an order of magnitude larger than our estimates of the maximum dust injection rates from AGB stars and core-collapse SNe, implying that dust growth by accretion in the ISM may be important in explaining the current IR emission in the MCs and other galaxies.

\acknowledgements{We thank Rick Arendt, Jon Slavin, Xander Tielens, Eric Pellegrini, and the participants of the MEGASage 7 collaboration meeting for the useful discussions that helped improve the paper. This work was supported by the NASA ADAP program NNH12ZDA001N-ADAP (proposal ID: 12-ADAP12-0145).}

%\bibliography{Astro_BIB.bib}
\bibliographystyle{apj}

\newpage

\begin{deluxetable*}{llcccccccccc}
\tablecolumns{12} \tablewidth{0pc} \tablecaption{\label{snrs_lmc}Parameters for the LMC SNRs}
%\tablehead{
%\colhead{PARAMETER}   &   \colhead{Region 1}   &   \colhead{Region 2}   &   \colhead{Region 3}   &   \colhead{Region 4}   &   \colhead{Region 5}   &   \colhead{Region 6}   &   \colhead{Region 7}   &   \colhead{Region 8}}
%\startdata
\tablehead{
  \colhead{SNR}   &    \colhead{Name}   &   \multicolumn{2}{c}{Position}   &   \colhead{D}    &   \colhead{$N_H$}   &    \colhead{D2G}   &   \colhead{$n_0$}   &    \multicolumn{2}{c}{$m_g(M_{\odot})$}   &    \multicolumn{2}{c}{$m_d(M_{\odot})$}   \\
    \colhead{}   &    \colhead{}   &   \colhead{RA}   &   \colhead{Dec.}   &   \colhead{($\arcsec$)}    &   \colhead{$\rm (10^{21} cm^{-2})$}   &    \colhead{($10^{-3}$)}   &   \colhead{$\rm (cm^{-3})$}   &   \colhead{Si}   &   \colhead{C}   &   \colhead{Si}   &   \colhead{C}   
    }
\startdata
J0448.4-6660	 & 	\nodata	 & 	04h 48m 22s	 & 	-66d 59m 52s	 & 	220	 & 	1.59	 & 	2.07	 & 	1.28	 & 	1889	 & 	1411	 & 	2.9	 & 	0.7		\\
J0449.3-6920	 & 	\nodata	 & 	04h 49m 20s	 & 	-69d 20m 20s	 & 	133	 & 	2.98	 & 	5.85	 & 	2.41	 & 	1764	 & 	1316	 & 	7.7	 & 	2.0	\\
J0449.7-6922	 & 	B0449-693	 & 	04h 49m 40s	 & 	-69d 21m 49s	 & 	120	 & 	3.30	 & 	5.92	 & 	2.67	 & 	1744	 & 	1301	 & 	7.7	 & 	2.0	\\
J0450.2-6922	 & 	B0450-6927	 & 	04h 50m 15s	 & 	-69d 22m 12s	 & 	210	 & 	2.77	 & 	5.78	 & 	2.24	 & 	1778	 & 	1326	 & 	7.6	 & 	2.0		\\
J0450.4-7050	 & 	B0450-709	 & 	04h 50m 27s	 & 	-70d 50m 15s	 & 	357	 & 	1.33	 & 	2.87	 & 	1.08	 & 	1925	 & 	1439	 & 	4.1	 & 	1.1		\\
J0453.2-6655	 & 	N4	 & 	04h 53m 14s	 & 	-66d 55m 13s	 & 	252	 & 	2.07	 & 	5.03	 & 	1.68	 & 	1835	 & 	1370	 & 	6.9	 & 	1.8		\\
J0453.6-6829	 & 	B0453-685	 & 	04h 53m 38s	 & 	-68d 29m 27s	 & 	120	 & 	1.72	 & 	2.87	 & 	1.39	 & 	1873	 & 	1399	 & 	4.0	 & 	1.0		\\
J0453.9-7000	 & 	B0454-7005	 & 	04h 53m 52s	 & 	-70d 00m 13s	 & 	420	 & 	0.75	 & 	3.03	 & 	0.60	 & 	2047	 & 	1531	 & 	4.6	 & 	1.2		\\
J0454.6-6713	 & 	N9	 & 	04h 54m 33s	 & 	-67d 13m 13s	 & 	177	 & 	1.44	 & 	4.03	 & 	1.17	 & 	1909	 & 	1426	 & 	5.7	 & 	1.5	 \\
J0454.8-6626	 & 	N11L	 & 	04h 54m 49s	 & 	-66d 25m 32s	 & 	87	 & 	3.24	 & 	4.17	 & 	2.63	 & 	1748	 & 	1303	 & 	5.4	 & 	1.4		\\
J0455.6-6839	 & 	N86	 & 	04h 55m 37s	 & 	-68d 38m 47s	 & 	348	 & 	1.73	 & 	3.82	 & 	1.40	 & 	1872	 & 	1398	 & 	5.3	 & 	1.4	 	\\
J0459.9-7008	 & 	N186D	 & 	04h 59m 55s	 & 	-70d 07m 52s	 & 	150	 & 	0.99	 & 	4.98	 & 	0.80	 & 	1987	 & 	1486	 & 	7.3	 & 	1.9	 	\\
J0505.7-6753	 & 	DEM L71	 & 	05h 05m 42s	 & 	-67d 52m 39s	 & 	72	 & 	1.27	 & 	2.75	 & 	1.03	 & 	1934	 & 	1445	 & 	4.0	 & 	1.0	 	\\
J0505.9-6802	 & 	N23	 & 	05h 05m 55s	 & 	-68d 01m 47s	 & 	111	 & 	1.86	 & 	4.44	 & 	1.51	 & 	1857	 & 	1387	 & 	6.1	 & 	1.6		\\
J0506.1-6541	 & 	\nodata	 & 	05h 06m 05s	 & 	-65d 41m 08s	 & 	408	 & 	1.49	 & 	2.48	 & 	1.21	 & 	1902	 & 	1421	 & 	3.5	 & 	0.9		\\
J0506.8-7026	 & 	B0507-7029	 & 	05h 06m 50s	 & 	-70d 25m 53s	 & 	330	 & 	0.75	 & 	3.76	 & 	0.61	 & 	2046	 & 	1530	 & 	5.7	 & 	1.5	 	\\
J0508.8-6831	 & 	\nodata	 & 	05h 08m 49s	 & 	-68d 30m 41s	 & 	108	 & 	2.55	 & 	5.00	 & 	2.06	 & 	1795	 & 	1339	 & 	6.7	 & 	1.7	 	\\
J0509.0-6844	 & 	N103Bk	 & 	05h 08m 59s	 & 	-68d 43m 35s	 & 	28	 & 	3.17	 & 	7.81	 & 	2.57	 & 	1752	 & 	1307	 & 	10.2	 & 	2.6		\\
J0509.5-6731	 & 	B0509-67.5	 & 	05h 09m 31s	 & 	-67d 31m 17s	 & 	29	 & 	1.11	 & 	1.56	 & 	0.90	 & 	1963	 & 	1467	 & 	2.3	 & 	0.6		\\
J0511.2-6759	 & 	\nodata	 & 	05h 11m 11s	 & 	-67d 59m 07s	 & 	108	 & 	1.94	 & 	2.32	 & 	1.57	 & 	1849	 & 	1381	 & 	3.2	 & 	0.8	 	\\
J0513.2-6912	 & 	DEM L109	 & 	05h 13m 14s	 & 	-69d 12m 20s	 & 	215	 & 	2.23	 & 	4.61	 & 	1.81	 & 	1821	 & 	1359	 & 	6.2	 & 	1.6	 	\\
J0514.3-6840	 & 	\nodata	 & 	05h 14m 15s	 & 	-68d 40m 14s	 & 	218	 & 	2.04	 & 	2.32	 & 	1.65	 & 	1838	 & 	1372	 & 	3.2	 & 	0.8	 	\\
J0517.2-6759	 & 	\nodata	 & 	05h 17m 10s	 & 	-67d 59m 03s	 & 	270	 & 	2.09	 & 	3.91	 & 	1.69	 & 	1834	 & 	1369	 & 	5.3	 & 	1.4	 	\\
J0518.7-6939	 & 	N120	 & 	05h 18m 41s	 & 	-69d 39m 12s	 & 	134	 & 	2.08	 & 	6.41	 & 	1.68	 & 	1834	 & 	1370	 & 	8.7	 & 	2.3	 	\\
J0519.6-6902	 & 	B0519-690	 & 	05h 19m 35s	 & 	-69d 02m 09s	 & 	31	 & 	1.20	 & 	2.62	 & 	0.97	 & 	1946	 & 	1454	 & 	3.8	 & 	1.0	 	\\
J0519.7-6926	 & 	B0520-694	 & 	05h 19m 44s	 & 	-69d 26m 08s	 & 	174	 & 	1.84	 & 	3.11	 & 	1.49	 & 	1858	 & 	1388	 & 	4.3	 & 	1.1	 	\\
J0521.6-6543	 & 	\nodata	 & 	05h 21m 39s	 & 	-65d 43m 07s	 & 	90	 & 	0.66	 & 	7.09	 & 	0.53	 & 	2075	 & 	1552	 & 	10.9	 & 	2.8	 	\\
J0523.1-6753	 & 	N44	 & 	05h 23m 07s	 & 	-67d 53m 12s	 & 	228	 & 	3.20	 & 	6.21	 & 	2.60	 & 	1750	 & 	1305	 & 	8.1	 & 	2.1	 \\
J0524.3-6624	 & 	DEM L175a	 & 	05h 24m 20s	 & 	-66d 24m 23s	 & 	234	 & 	3.05	 & 	4.24	 & 	2.47	 & 	1760	 & 	1313	 & 	5.5	 & 	1.4	 	\\
J0525.1-6938	 & 	N132D	 & 	05h 25m 04s	 & 	-69d 38m 24s	 & 	114	 & 	1.74	 & 	8.26	 & 	1.41	 & 	1870	 & 	1397	 & 	11.5	 & 	3.0		\\
J0525.4-6559	 & 	N49B	 & 	05h 25m 25s	 & 	-65d 59m 19s	 & 	168	 & 	2.50	 & 	4.20	 & 	2.02	 & 	1798	 & 	1342	 & 	5.6	 & 	1.4	 	\\
J0526.0-6605	 & 	N49	 & 	05h 26m 00s	 & 	-66d 04m 57s	 & 	84	 & 	4.10	 & 	6.25	 & 	3.32	 & 	1703	 & 	1269	 & 	7.9	 & 	2.0	 	\\
J0527.6-6912	 & 	B0528-692	 & 	05h 27m 39s	 & 	-69d 12m 04s	 & 	147	 & 	0.83	 & 	3.57	 & 	0.67	 & 	2024	 & 	1514	 & 	5.4	 & 	1.4	 	\\
J0527.9-6550	 & 	DEM L204	 & 	05h 27m 54s	 & 	-65d 49m 38s	 & 	303	 & 	1.42	 & 	3.08	 & 	1.15	 & 	1911	 & 	1428	 & 	4.4	 & 	1.1	 	\\
J0527.9-6714	 & 	B0528-6716	 & 	05h 27m 56s	 & 	-67d 13m 40s	 & 	196	 & 	0.98	 & 	3.40	 & 	0.80	 & 	1988	 & 	1486	 & 	5.0	 & 	1.3		\\
J0528.1-7038	 & 	B0528-7038	 & 	05h 28m 03s	 & 	-70d 37m 40s	 & 	60	 & 	0.77	 & 	3.85	 & 	0.62	 & 	2041	 & 	1527	 & 	5.8	 & 	1.5	\\
J0528.3-6714	 & 	HP99498	 & 	05h 28m 20s	 & 	-67d 13m 40s	 & 	97	 & 	0.92	 & 	3.11	 & 	0.74	 & 	2002	 & 	1497	 & 	4.6	 & 	1.2		\\
J0529.1-6833	 & 	DEM L203	 & 	05h 29m 05s	 & 	-68d 32m 30s	 & 	667	 & 	1.90	 & 	5.71	 & 	1.54	 & 	1852	 & 	1383	 & 	7.9	 & 	2.0		\\
J0529.9-6701	 & 	DEM L214	 & 	05h 29m 51s	 & 	-67d 01m 05s	 & 	100	 & 	0.69	 & 	2.85	 & 	0.56	 & 	2064	 & 	1544	 & 	4.4	 & 	1.1		\\
J0530.7-7008	 & 	DEM L218	 & 	05h 30m 40s	 & 	-70d 07m 30s	 & 	213	 & 	1.23	 & 	1.85	 & 	1.00	 & 	1940	 & 	1450	 & 	2.7	 & 	0.7		\\
J0531.9-7100	 & 	N206	 & 	05h 31m 56s	 & 	-71d 00m 19s	 & 	192	 & 	1.94	 & 	4.95	 & 	1.57	 & 	1848	 & 	1380	 & 	6.8	 & 	1.8		\\
J0532.5-6732	 & 	B0532-675	 & 	05h 32m 30s	 & 	-67d 31m 33s	 & 	252	 & 	1.61	 & 	5.52	 & 	1.30	 & 	1886	 & 	1409	 & 	7.7	 & 	2.0		\\
J0534.0-6955	 & 	B0534-699	 & 	05h 34m 02s	 & 	-69d 55m 03s	 & 	114	 & 	2.30	 & 	2.69	 & 	1.86	 & 	1815	 & 	1355	 & 	3.6	 & 	0.9		\\
J0534.3-7033	 & 	DEM L238	 & 	05h 34m 18s	 & 	-70d 33m 26s	 & 	180	 & 	1.57	 & 	1.31	 & 	1.27	 & 	1891	 & 	1413	 & 	1.8	 & 	0.5		\\
J0535.5-6916	 & 	SNR1987A	 & 	05h 35m 28s	 & 	-69d 16m 11s	 & 	2	 & 	1.86	 & 	4.15	 & 	1.51	 & 	1856	 & 	1386	 & 	5.7	 & 	1.5		\\
J0535.7-6602	 & 	N63A	 & 	05h 35m 44s	 & 	-66d 02m 14s	 & 	66	 & 	1.24	 & 	9.62	 & 	1.00	 & 	1939	 & 	1449	 & 	13.9	 & 	3.6		\\
J0535.8-6918	 & 	Honeycomb	 & 	05h 35m 46s	 & 	-69d 18m 02s	 & 	102	 & 	2.28	 & 	5.03	 & 	1.84	 & 	1817	 & 	1356	 & 	6.8	 & 	1.8		\\
J0536.1-6735	 & 	DEM L241	 & 	05h 36m 03s	 & 	-67d 34m 36s	 & 	135	 & 	3.21	 & 	6.62	 & 	2.60	 & 	1750	 & 	1305	 & 	8.6	 & 	2.2		\\
J0536.1-7039	 & 	DEM L249	 & 	05h 36m 07s	 & 	-70d 38m 37s	 & 	180	 & 	2.54	 & 	2.29	 & 	2.06	 & 	1795	 & 	1340	 & 	3.1	 & 	0.8	\\
J0536.2-6912	 & 	B0536-6914	 & 	05h 36m 09s	 & 	-69d 11m 53s	 & 	480	 & 	3.28	 & 	5.68	 & 	2.66	 & 	1745	 & 	1301	 & 	7.4	 & 	1.9		\\
J0537.4-6628	 & 	DEM L256	 & 	05h 37m 27s	 & 	-66d 27m 50s	 & 	204	 & 	1.92	 & 	6.94	 & 	1.55	 & 	1850	 & 	1382	 & 	9.5	 & 	2.5	 	\\
J0537.6-6920	 & 	B0538-6922	 & 	05h 37m 37s	 & 	-69d 20m 23s	 & 	169	 & 	3.43	 & 	4.69	 & 	2.78	 & 	1737	 & 	1295	 & 	6.1	 & 	1.6		\\
J0537.8-6910	 & 	N157B	 & 	05h 37m 46s	 & 	-69d 10m 28s	 & 	102	 & 	5.35	 & 	9.17	 & 	4.33	 & 	1654	 & 	1232	 & 	11.3	 & 	2.9	 	\\
J0538.2-6922	 & 	0538-693	 & 	05h 38m 14s	 & 	-69d 21m 36s	 & 	169	 & 	3.67	 & 	5.00	 & 	2.97	 & 	1724	 & 	1285	 & 	6.4	 & 	1.7	 	\\
J0540.0-6944	 & 	N159	 & 	05h 39m 59s	 & 	-69d 44m 02s	 & 	78	 & 	6.33	 & 	10.4	 & 	5.12	 & 	1622	 & 	1207	 & 	12.5	 & 	3.2	 	\\
J0540.2-6920	 & 	B0540-693	 & 	05h 40m 11s	 & 	-69d 19m 55s	 & 	60	 & 	4.72	 & 	4.78	 & 	3.82	 & 	1676	 & 	1249	 & 	6.0	 & 	1.5		\\
J0543.1-6858	 & 	DEM L299	 & 	05h 43m 08s	 & 	-68d 58m 18s	 & 	318	 & 	4.73	 & 	3.39	 & 	3.83	 & 	1676	 & 	1249	 & 	4.2	 & 	1.1	 	\\
J0547.0-6943	 & 	DEM L316B	 & 	05h 46m 59s	 & 	-69d 42m 50s	 & 	84	 & 	5.88	 & 	4.85	 & 	4.76	 & 	1635	 & 	1217	 & 	5.9	 & 	1.5		\\
J0547.4-6941	 & 	DEM L316A	 & 	05h 47m 22s	 & 	-69d 41m 26s	 & 	56	 & 	6.20	 & 	4.44	 & 	5.02	 & 	1626	 & 	1210	 & 	5.4	 & 	1.4	 	\\
J0547.8-7025	 & 	B0548-704	 & 	05h 47m 49s	 & 	-70d 24m 54s	 & 	102	 & 	3.45	 & 	2.54	 & 	2.80	 & 	1735	 & 	1294	 & 	3.3	 & 	0.8		\\
J0550.5-6823	 & 	\nodata	 & 	05h 50m 30s	 & 	-68d 22m 40s	 & 	312	 & 	2.62	 & 	3.10	 & 	2.12	 & 	1789	 & 	1335	 & 	4.1	 & 	1.1	
\enddata
\tablecomments{Listed values for the density ($n_0$), effective swept-up gas mass ($m_g$), and destroyed dust mass ($m_d$) assume a gas disk thickness of 0.4 kpc. Of the total D2G mass ratio, 74\% is attributed to silicates, and 26\% to carbon dust.}
\end{deluxetable*}

\begin{deluxetable*}{llcccccccccc}
\tablecolumns{12} \tablewidth{0pc} \tablecaption{\label{snrs_smc}Parameters for the SMC SNRs}
%\tablehead{

%\tablehead{
%\colhead{PARAMETER}   &   \colhead{Region 1}   &   \colhead{Region 2}   &   \colhead{Region 3}   &   \colhead{Region 4}   &   \colhead{Region 5}   &   \colhead{Region 6}   &   \colhead{Region 7}   &   \colhead{Region 8}}
%\startdata
\tablehead{
  \colhead{SNR}   &    \colhead{Name}   &   \multicolumn{2}{c}{Position}   &   \colhead{D}    &   \colhead{$N_H$}   &    \colhead{D2G}   &   \colhead{$n_0$}   &    \multicolumn{2}{c}{$m_g(M_{\odot})$}   &    \multicolumn{2}{c}{$m_d(M_{\odot})$}    \\
    \colhead{}   &    \colhead{}   &   \colhead{RA}   &   \colhead{Dec.}   &   \colhead{($\arcsec$)}    &   \colhead{$\rm (10^{21} cm^{-2})$}   &    \colhead{($10^{-3}$)}   &   \colhead{$\rm (cm^{-3})$}   &   \colhead{Si}   &   \colhead{C}   &   \colhead{Si}   &   \colhead{C}   
    }
\startdata
J0040.9-7337	 & 	DEM S5	 & 	00h 40m 55s	 & 	-73d 36m 55s	 & 	121	 & 	3.17	 & 	0.50	 & 	0.51	 & 	2082	 & 	1558	 & 	0.8	 & 	0.2	 	\\
J0046.6-7309	 & 	DEM S32	 & 	00h 46m 39s	 & 	-73d 08m 39s	 & 	136	 & 	10.30	 & 	1.61	 & 	1.67	 & 	1836	 & 	1371	 & 	2.4	 & 	0.4		\\
J0047.2-7308	 & 	IKT2	 & 	00h 47m 12s	 & 	-73d 08m 26s	 & 	66	 & 	10.89	 & 	1.76	 & 	1.76	 & 	1825	 & 	1363	 & 	2.6	 & 	0.5	 	\\
J0047.5-7306	 & 	B0045-733	 & 	00h 47m 29s	 & 	-73d 06m 01s	 & 	180	 & 	10.58	 & 	1.80	 & 	1.71	 & 	1831	 & 	1367	 & 	2.7	 & 	0.5	 	\\
J0047.7-7310	 & 	HFPK419	 & 	00h 47m 41s	 & 	-73d 09m 30s	 & 	90	 & 	10.66	 & 	1.96	 & 	1.73	 & 	1829	 & 	1366	 & 	2.9	 & 	0.5		\\
J0047.8-7317	 & 	NS21	 & 	00h 47m 48s	 & 	-73d 17m 27s	 & 	76	 & 	9.66	 & 	2.42	 & 	1.56	 & 	1849	 & 	1381	 & 	3.6	 & 	0.7	 \\
J0048.1-7309	 & 	NS19	 & 	00h 48m 06s	 & 	-73d 08m 43s	 & 	79	 & 	11.65	 & 	2.25	 & 	1.89	 & 	1812	 & 	1352	 & 	3.3	 & 	0.6	 	\\
J0048.4-7319	 & 	IKT4	 & 	00h 48m 25s	 & 	-73d 19m 24s	 & 	84	 & 	8.98	 & 	1.87	 & 	1.46	 & 	1864	 & 	1392	 & 	2.8	 & 	0.5	 	\\
J0049.1-7314	 & 	IKT5	 & 	00h 49m 07s	 & 	-73d 14m 05s	 & 	116	 & 	9.11	 & 	1.64	 & 	1.48	 & 	1861	 & 	1390	 & 	2.5	 & 	0.5		\\
J0051.1-7321	 & 	IKT6	 & 	00h 51m 07s	 & 	-73d 21m 26s	 & 	144	 & 	6.98	 & 	1.29	 & 	1.13	 & 	1915	 & 	1431	 & 	2.0	 & 	0.4		\\
J0051.9-7310	 & 	IKT7	 & 	00h 51m 54s	 & 	-73d 10m 24s	 & 	97	 & 	7.23	 & 	1.43	 & 	1.17	 & 	1908	 & 	1425	 & 	2.2	 & 	0.4		\\
J0052.6-7238	 & 	B0050-728	 & 	00h 52m 33s	 & 	-72d 37m 35s	 & 	323	 & 	4.89	 & 	0.96	 & 	0.79	 & 	1989	 & 	1487	 & 	1.5	 & 	0.3	 	\\
J0058.3-7218	 & 	IKT16	 & 	00h 58m 16s	 & 	-72d 18m 05s	 & 	200	 & 	5.69	 & 	1.41	 & 	0.92	 & 	1957	 & 	1463	 & 	2.2	 & 	0.4	 	\\
J0059.4-7210	 & 	IKT18	 & 	00h 59m 25s	 & 	-72d 10m 10s	 & 	158	 & 	3.92	 & 	2.26	 & 	0.64	 & 	2036	 & 	1523	 & 	3.7	 & 	0.7		\\
J0100.3-7134	 & 	DEM S108	 & 	01h 00m 21s	 & 	-71h 33m 40s	 & 	149	 & 	3.06	 & 	1.55	 & 	0.50	 & 	2090	 & 	1563	 & 	2.6	 & 	0.5		\\
J0103.2-7209	 & 	IKT21	 & 	01h 03m 13s	 & 	-72d 08m 59s	 & 	62	 & 	5.48	 & 	1.48	 & 	0.89	 & 	1965	 & 	1469	 & 	2.3	 & 	0.4		\\
J0103.5-7247	 & 	HFPK334	 & 	01h 03m 30s	 & 	-72d 47m 20s	 & 	86	 & 	4.35	 & 	0.74	 & 	0.70	 & 	2014	 & 	1506	 & 	1.2	 & 	0.2		\\
J0104.0-7202	 & 	B0102-7219	 & 	01h 04m 02s	 & 	-72d 01m 48s	 & 	44	 & 	5.24	 & 	2.11	 & 	0.85	 & 	1975	 & 	1476	 & 	3.3	 & 	0.6		\\
J0105.1-7223	 & 	IKT23	 & 	01h 05m 04s	 & 	-72d 22m 56s	 & 	170	 & 	4.13	 & 	0.91	 & 	0.67	 & 	2025	 & 	1515	 & 	1.5	 & 	0.3		\\
J0105.4-7209d	 & 	DEM S128	 & 	01h 05m 23s	 & 	-72d 09m 26s	 & 	124	 & 	4.66	 & 	1.35	 & 	0.75	 & 	1999	 & 	1495	 & 	2.2	 & 	0.4	\\
J0105.6-7204	 & 	DEM S130	 & 	01h 05m 39s	 & 	-72d 03m 41s	 & 	79	 & 	5.61	 & 	1.76	 & 	0.91	 & 	1960	 & 	1465	 & 	2.8	 & 	0.5	\\
J0106.2-7205	 & 	IKT25	 & 	01h 06m 14s	 & 	-72d 05m 18s	 & 	110	 & 	5.58	 & 	1.43	 & 	0.90	 & 	1961	 & 	1466	 & 	2.3	 & 	0.4		\\
J0114.0-7317	 & 	N83C	 & 	01h 14m 00s	 & 	-73d 17m 08s	 & 	17	 & 	5.50	 & 	5.41	 & 	0.89	 & 	1965	 & 	1468	 & 	8.5	 & 	1.6		
\enddata
\tablecomments{Listed values for the density ($n_0$), effective swept-up gas mass ($m_g$),  and destroyed dust mass ($m_d$) assume a gas disk thickness of 2.0 kpc. Of the total D2G mass ratio, 80\% is attributed to silicates, and 20\% to carbon dust.}
\end{deluxetable*}

\begin{deluxetable*}{ccccccccccc}
\tablecolumns{11} \tablewidth{0pc} \tablecaption{\label{snrs_avg}AVERAGE VALUES VS. DISK THICKNESS}
%\tablehead{
%\colhead{PARAMETER} & \colhead{Region 1} & \colhead{Region 2} & \colhead{Region 3} & \colhead{Region 4} & \colhead{Region 5} & \colhead{Region 6} & \colhead{Region 7} & \colhead{Region 8}}
%\startdata
\tablehead{
  \colhead{d} &  \colhead{$\overline{n}_0$} & \colhead{$\overline{R}_{SN}$} &\multicolumn{2}{c}{\mg$(M_{\odot})$} & \multicolumn{2}{c}{$\mde(M_{\odot})$}  & \multicolumn{2}{c}{$\tau_d$(Myr)} & \multicolumn{2}{c}{$dM_d(t)/dt\: (10^{-3}\:M_{\odot}/yr)$} \\
    \colhead{(kpc)} &  \colhead{$\rm (cm^{-3})$} & \colhead{$(\rm 10^{-3}\: yr^{-1})$} & \colhead{Si} & \colhead{C} & \colhead{Si} & \colhead{C} & \colhead{Si} & \colhead{C}  & \colhead{Si} & \colhead{C} 
        }
\startdata
\cutinhead{LMC}
%0.05	&	15.3	 $\pm$ 	8.9	&	12.3	 $\pm$ 	4.71	&	1457	 $\pm$ 	100	&	1078	 $\pm$ 	78	&	4.8	 $\pm$ 	2.0	&	1.2	 $\pm$ 	0.5	&	9	 $\pm$ 	5	&	12	 $\pm$ 	7	&	59.3	 $\pm$ 	33.6	&	15.2	 $\pm$ 	8.6	\\
0.1	&	7.6	 $\pm$ 	4.4	&	8.29	 $\pm$ 	3.17	&	1580	 $\pm$ 	103	&	1174	 $\pm$ 	80	&	5.2	 $\pm$ 	2.2	&	1.3	 $\pm$ 	0.6	&	12	 $\pm$ 	7	&	16	 $\pm$ 	9	&	43.3	 $\pm$ 	25	&	11.1	 $\pm$ 	6.3	\\
0.2	&	3.8	 $\pm$ 	2.2	&	5.58	 $\pm$ 	2.13	&	1709	 $\pm$ 	108	&	1273	 $\pm$ 	83	&	5.7	 $\pm$ 	2.4	&	1.5	 $\pm$ 	0.6	&	16	 $\pm$ 	9	&	22	 $\pm$ 	13	&	31.5	 $\pm$ 	18	&	8.1	 $\pm$ 	4.6	\\
0.4	&	1.9	 $\pm$ 	1.1	&	3.75	 $\pm$ 	1.44	&	1843	 $\pm$ 	114	&	1376	 $\pm$ 	87	&	6.1	 $\pm$ 	2.6	&	1.6	 $\pm$ 	0.7	&	22	 $\pm$ 	13	&	30	 $\pm$ 	17	&	22.9	 $\pm$ 	13	&	5.9	 $\pm$ 	3.4	\\
0.7	&	1.1	 $\pm$ 	0.6	&	2.73	 $\pm$ 	1.04	&	1957	 $\pm$ 	118	&	1463	 $\pm$ 	90	&	6.5	 $\pm$ 	2.7	&	1.7	 $\pm$ 	0.7	&	29	 $\pm$ 	17	&	39	 $\pm$ 	22	&	17.7	 $\pm$ 	10	&	4.6	 $\pm$ 	2.6	\\
%1.0	&	0.76	 $\pm$ 	0.44	&	2.22	 $\pm$ 	0.85	&	2032	 $\pm$ 	121	&	1519	 $\pm$ 	91	&	6.7	 $\pm$ 	2.8	&	1.7	 $\pm$ 	0.7	&	34	 $\pm$ 	20	&	46	 $\pm$ 	26	&	15.0	 $\pm$ 	8.5	&	3.9	 $\pm$ 	2.2	\\
1.6	&	0.5	 $\pm$ 	0.3	&	1.70	 $\pm$ 	0.65	&	2133	 $\pm$ 	123	&	1594	 $\pm$ 	91	&	7.1	 $\pm$ 	3.0	&	1.8	 $\pm$ 	0.8	&	43	 $\pm$ 	25	&	57	 $\pm$ 	33	&	12.0	 $\pm$ 	6.8	&	3.1	 $\pm$ 	1.8	\\
%2.0	&	0.38	 $\pm$ 	0.22	&	1.50	 $\pm$ 	0.57	&	2181	 $\pm$ 	123	&	1630	 $\pm$ 	91	&	7.2	 $\pm$ 	3.0	&	1.9	 $\pm$ 	0.8	&	48	 $\pm$ 	27	&	64	 $\pm$ 	37	&	10.8	 $\pm$ 	6.1	&	2.8	 $\pm$ 	1.6	\\
%3.0	&	0.25	 $\pm$ 	0.15	&	1.19	 $\pm$ 	0.45	&	2269	 $\pm$ 	124	&	1694	 $\pm$ 	90	&	7.5	 $\pm$ 	3.2	&	1.9	 $\pm$ 	0.8	&	58	 $\pm$ 	33	&	77	 $\pm$ 	44	&	8.9	 $\pm$ 	5.1	&	2.3	 $\pm$ 	1.3	\\
%4.0	&	0.19	 $\pm$ 	0.11	&	1.01	 $\pm$ 	0.39	&	2331	 $\pm$ 	124	&	1739	 $\pm$ 	88	&	7.7	 $\pm$ 	3.3	&	2.0	 $\pm$ 	0.8	&	66	 $\pm$ 	38	&	89	 $\pm$ 	51	&	7.8	 $\pm$ 	4.4	&	2.0	 $\pm$ 	1.1	\\
%5.0	&	0.15	 $\pm$ 	0.09	&	0.89	 $\pm$ 	0.34	&	2380	 $\pm$ 	124	&	1774	 $\pm$ 	87	&	7.9	 $\pm$ 	3.3	&	2.0	 $\pm$ 	0.9	&	74	 $\pm$ 	42	&	99	 $\pm$ 	57	&	7.0	 $\pm$ 	4.0	&	1.8	 $\pm$ 	1.0	\\
%6.0	&	0.13	 $\pm$ 	0.07	&	0.80	 $\pm$ 	0.31	&	2419	 $\pm$ 	123	&	1801	 $\pm$ 	86	&	8.0	 $\pm$ 	3.4	&	2.1	 $\pm$ 	0.9	&	80	 $\pm$ 	46	&	108	 $\pm$ 	62	&	6.4	 $\pm$ 	3.6	&	1.6	 $\pm$ 	0.9	\\
\cutinhead{SMC}																																									
%0.05	&	44.3	 $\pm$ 	17.9	&	9.15	 $\pm$ 	2.26	&	1259	 $\pm$ 	70	&	923	 $\pm$ 	56	&	1.7	 $\pm$ 	1.0	&	0.32	 $\pm$ 	0.17	&	10	 $\pm$ 	6	&	14	 $\pm$ 	8	&	16.0	 $\pm$ 	9.6	&	2.91	 $\pm$ 	1.75	\\
%0.1	&	22.2	 $\pm$ 	8.9	&	6.16	 $\pm$ 	1.52	&	1378	 $\pm$ 	71	&	1017	 $\pm$ 	56	&	1.9	 $\pm$ 	1.0	&	0.35	 $\pm$ 	0.19	&	14	 $\pm$ 	8	&	18	 $\pm$ 	11	&	11.8	 $\pm$ 	7.0	&	2.16	 $\pm$ 	1.30	\\
%0.2	&	11.1	 $\pm$ 	4.5	&	4.14	 $\pm$ 	1.03	&	1499	 $\pm$ 	73	&	1112	 $\pm$ 	57	&	2.1	 $\pm$ 	1.1	&	0.38	 $\pm$ 	0.21	&	19	 $\pm$ 	11	&	25	 $\pm$ 	15	&	8.6	 $\pm$ 	5.2	&	1.59	 $\pm$ 	0.95	\\
%0.4	&	5.5	 $\pm$ 	2.2	&	2.79	 $\pm$ 	0.69	&	1624	 $\pm$ 	76	&	1208	 $\pm$ 	58	&	2.3	 $\pm$ 	1.2	&	0.42	 $\pm$ 	0.23	&	25	 $\pm$ 	15	&	34	 $\pm$ 	21	&	6.3	 $\pm$ 	3.8	&	1.16	 $\pm$ 	0.70	\\
0.7	&	3.2	 $\pm$ 	1.3	&	2.03	 $\pm$ 	0.50	&	1728	 $\pm$ 	79	&	1289	 $\pm$ 	60	&	2.4	 $\pm$ 	1.3	&	0.45	 $\pm$ 	0.24	&	33	 $\pm$ 	20	&	44	 $\pm$ 	27	&	4.9	 $\pm$ 	2.9	&	0.90	 $\pm$ 	0.54	\\
1.0	&	2.2	 $\pm$ 	0.9	&	1.65	 $\pm$ 	0.41	&	1797	 $\pm$ 	81	&	1341	 $\pm$ 	62	&	2.5	 $\pm$ 	1.4	&	0.46	 $\pm$ 	0.25	&	39	 $\pm$ 	24	&	52	 $\pm$ 	32	&	4.1	 $\pm$ 	2.5	&	0.77	 $\pm$ 	0.46	\\
%1.6	&	1.4	 $\pm$ 	0.56	&	1.26	 $\pm$ 	0.31	&	1891	 $\pm$ 	84	&	1413	 $\pm$ 	64	&	2.6	 $\pm$ 	1.4	&	0.49	 $\pm$ 	0.27	&	48	 $\pm$ 	29	&	65	 $\pm$ 	39	&	3.3	 $\pm$ 	2.0	&	0.62	 $\pm$ 	0.37	\\
2.0	&	1.1	 $\pm$ 	0.5	&	1.11	 $\pm$ 	0.27	&	1937	 $\pm$ 	85	&	1447	 $\pm$ 	65	&	2.7	 $\pm$ 	1.5	&	0.50	 $\pm$ 	0.27	&	54	 $\pm$ 	32	&	72	 $\pm$ 	43	&	3.0	 $\pm$ 	1.8	&	0.56	 $\pm$ 	0.33	\\
%3.0	&	0.74	 $\pm$ 	0.30	&	0.88	 $\pm$ 	0.22	&	2022	 $\pm$ 	88	&	1512	 $\pm$ 	66	&	2.8	 $\pm$ 	1.5	&	0.52	 $\pm$ 	0.28	&	65	 $\pm$ 	39	&	87	 $\pm$ 	52	&	2.5	 $\pm$ 	1.5	&	0.46	 $\pm$ 	0.28	\\
4.0	&	0.6	 $\pm$ 	0.2	&	0.75	 $\pm$ 	0.19	&	2083	 $\pm$ 	89	&	1558	 $\pm$ 	67	&	2.9	 $\pm$ 	1.6	&	0.54	 $\pm$ 	0.29	&	74	 $\pm$ 	45	&	99	 $\pm$ 	60	&	2.2	 $\pm$ 	1.3	&	0.40	 $\pm$ 	0.24	\\
5.0	&	0.4	 $\pm$ 	0.2	&	0.66	 $\pm$ 	0.16	&	2132	 $\pm$ 	89	&	1594	 $\pm$ 	67	&	3.0	 $\pm$ 	1.6	&	0.55	 $\pm$ 	0.30	&	82	 $\pm$ 	50	&	110	 $\pm$ 	67	&	1.9	 $\pm$ 	1.2	&	0.36	 $\pm$ 	0.22	\\
%6.0	&	0.37	 $\pm$ 	0.15	&	0.59	 $\pm$ 	0.15	&	2171	 $\pm$ 	90	&	1623	 $\pm$ 	66	&	3.0	 $\pm$ 	1.6	&	0.56	 $\pm$ 	0.31	&	90	 $\pm$ 	54	&	120	 $\pm$ 	73	&	1.8	 $\pm$ 	1.1	&	0.33	 $\pm$ 	0.20	
\enddata
\tablecomments{The uncertainties in $n_0$, \mg\ and \md\  represent the standard deviation, uncertainty on $R_{SNR}$ reflects the spread in visibility times $\tau_v$ for all SNRs in the sample. The uncertainties were propagated accordingly for the rest of the parameters, based on equations is \S\ref{equations}.}
\end{deluxetable*}

\begin{deluxetable}{lcccccc}
\tablecolumns{7} \tablewidth{0pc} \tablecaption{\label{params}DERIVED PARAMETERS}
%\tablehead{
%\colhead{PARAMETER} & \colhead{Region 1} & \colhead{Region 2} & \colhead{Region 3} & \colhead{Region 4} & \colhead{Region 5} & \colhead{Region 6} & \colhead{Region 7} & \colhead{Region 8}}
%\startdata
\tablehead{
  \colhead{Parameter} &   \colhead{Density} & \colhead{Metallicity}  & \multicolumn{2}{c}{LMC}  & \multicolumn{2}{c}{SMC}  \\
    \colhead{} & \colhead{Dependence} & \colhead{Dependence} & \colhead{Silicates} & \colhead{Carbon} & \colhead{Silicates} & \colhead{Carbon} 
        }
\startdata
\\
 $\overline{n}_0$ around SNRs ($\rm cm^{-3}$) & \nodata & \nodata &\multicolumn{2}{c}{$1.9\pm1.1$}  & \multicolumn{2}{c}{$1.1\pm0.5$}  \\
\\
$\overline{D2G}$ around SNRs ($10^{-3}$) & \nodata & \nodata & \multicolumn{2}{c}{$4.5\pm2.0$}  & \multicolumn{2}{c}{$1.7\pm0.9$}  \\
\\
$\overline{D2G}$ global ($10^{-3}$)& \nodata & \nodata & \multicolumn{2}{c}{$1.8$}  & \multicolumn{2}{c}{$0.8$}  \\
\\
$\overline{\tau}_{vis}$ (kyr) & $n_0^{-4/7}$  &  $\zeta_m^{-5/14}$  & \multicolumn{2}{c}{16.3} &  \multicolumn{2}{c}{20.7} \\
\\
$\overline{R}_{SN}$ ($10^{-3}\: \rm M_{\odot}/yr$) & $n_0^{4/7}$ & $\zeta_m^{5/14}$ & \multicolumn{2}{c}{3.75 $\pm$ 1.44 } &  \multicolumn{2}{c}{1.11 $\pm$ 0.27} \\
 \\
 $M_d$ ($10^5\: \rm M_{\odot}$) & \nodata & \nodata & $5.2\pm0.4$ & $1.8\pm0.1$ & $1.6\pm0.2$ & $0.40\pm0.04$ \\
 \\
 $T_d$ (K) & \nodata  & \nodata & 22.4 $\pm$ 0.4 & 26.9 $\pm$ 0.5 & 19.0 $\pm$ 0.5 & 22.8 $\pm$ 0.6 \\
\\
\mg $(\rm M_{\odot})$ & $n_0^{-0.107}$ & $\zeta_m^{-0.15}$ & 1840 $\pm$ 110 & 1380 $\pm$ 80 & 1940 $\pm$ 90 & 1450 $\pm$ 70 \\
\\
$\mde (\rm M_{\odot})$ & $n_0^{-0.107}$  & $\zeta_m^{-0.15}$ & 6.1 $\pm$ 2.6 & 1.6 $\pm$ 0.7 & 2.7 $\pm$ 1.5 & 0.57 $\pm$ 0.27  \\
\\
$\tau_d$ (Myr) & $n_0^{-0.464}$ & $\zeta_m^{-0.207}$ & 22 $\pm$ 13 & 30 $\pm$ 17 & 54 $\pm$ 32 & 72 $\pm$ 43 \\
\\
$dM_d(t)/dt\: (10^{-3}\rm M_{\odot}/yr)$ & $n_0^{0.464}$ & $\zeta_m^{\: 0.207}$ &  $23\pm13$ &$5.9\pm3.4$  &  $3.0\pm1.8$  &  $0.56\pm0.33$ \\
\enddata
\tablecomments{The assumed nominal disk thickness values for the LMC and SMC are 0.4 kpc and 2.0 kpc, respectively, based on estimates of \citet{kim99} and \citet{stanimirovic04}. The assumed metallicity factor $\zeta_m$ is 0.3. The dependence of the derived parameters on the ambient gas density is indicated in the second column. The disk thickness dependence for each parameter is just the inverse of the density dependence ($n_0 \propto d^{-1}$). We note that density and metallicity dependence for $\tau_d$ and  $dM_d(t)/dt$ is specific to analyses of galaxies with complete samples of resolved SNRs, since it involves the density and metallicity dependent  $R_{SN}$ and $\tau_{vis}$ (see Section \ref{application})}
\end{deluxetable}

\begin{deluxetable}{cccccc}
\tablecolumns{6} \tablewidth{0pc} \tablecaption{\label{rates}DUST INJECTION AND DESTRUCTION RATES \\ $dM_d(t)/dt\: (10^{-6}\:M_{\odot}/yr)$}
%\tablehead{
%\colhead{PARAMETER} & \colhead{Region 1} & \colhead{Region 2} & \colhead{Region 3} & \colhead{Region 4} & \colhead{Region 5} & \colhead{Region 6} & \colhead{Region 7} & \colhead{Region 8}}
%\startdata
\tablehead{
  \colhead{Source} &  \multicolumn{2}{c}{LMC}  & \multicolumn{2}{c}{SMC} &  \colhead{Reference} \\
    \colhead{} &  \colhead{Silicates} & \colhead{Carbon} & \colhead{Silicates} & \colhead{Carbon} & \colhead{}
        }
\startdata
\sidehead{AGB Stars} 
& 0.95--5.5 & 9.5--13.6 & 0.08 & 0.75 & Observed* \\
\\
& 31 & 108 & 21 & 85   & This work \\
\\
\hline
\sidehead{Core Collapse SNe} 
%Observation & \multicolumn{4}{c}{} & \\
 & $1.3\times10^3$ & $3.9\times10^2$ & $3.9\times10^2$ & $1.2\times10^2$ & This work \\
% & \multicolumn{2}{c}{$<6\times10^2$} & \multicolumn{2}{c}{$<3\times10^2$}  & \citet{kozasa09} \\
 \\
 \hline
\sidehead{Destruction by SNe} 
& $-2.3\times10^4$ & $-5.9\times10^3$ & $-3.0\times10^3$ & $-5.6\times10^2$  & This work \\
\enddata
\tablecomments{*References for observational measurements of dust injection by AGB stars for the LMC: \citet{srinivasan09}, \citet{boyer12}, \citet{riebel12}, and SMC: \citet{boyer12}. The dust injection rates from AGB stars and CCSNe estimated in this work assume a 100\% grain condensation efficiency, and do not include dust destruction by the SN reverse shock. They therefore represent absolute upper limits on the injected dust mass. 
 }
\end{deluxetable}

\end{document}